\documentclass [superscriptaddress,reprint,amsmath,amssymb,showkeys]{revtex4-2}
\setcitestyle{super}
\usepackage{graphicx}
\usepackage{dcolumn}
\usepackage{bm}
\usepackage{color}
\usepackage{float}
\usepackage[utf8]{inputenc}
\usepackage[mathlines]{lineno}
\usepackage{multirow}
\usepackage{amsmath}
\usepackage{xr-hyper}
\usepackage{hyperref}
\usepackage{soul,xcolor}
\usepackage{natbib}
\usepackage{amsmath} 
\usepackage{xcolor}
\usepackage{xr}
\soulregister\cite7
\soulregister\ref7
\setstcolor{blue}
\newcommand{\dBO}{$\delta$--Bi$_2$O$_3$}
\newcommand{\aBO}{$\alpha$--Bi$_2$O$_3$}
\newcommand{\BO}{Bi$_2$O$_3$}

\newcommand{\wn}{cm$^{-1}$}

\newcommand{\degs}{$^\circ$}

\newcommand{\doHMN}[2]{%
  \begingroup\lccode`~=`#1
  \lowercase{\endgroup\let~}#2%
  \mathcode`#1="8000
}

\makeatletter
\newcommand*{\addFileDependency}[1]{
  \typeout{(#1)}
  \@addtofilelist{#1}
  \IfFileExists{#1}{}{\typeout{No file #1.}}
}
\makeatother

\begin{document}
\title{Dynamic disorder in Bi sub-lattice of \dBO}

\author{Rituraj Sharma}
\affiliation{Department of Chemical and Biological Physics, Weizmann Institute of Science, Rehovot 76100, Israel}
\author{Nimrod Benshalom}
\affiliation{Department of Chemical and Biological Physics, Weizmann Institute of Science, Rehovot 76100, Israel}
\author{Maor Asher}
\affiliation{Department of Chemical and Biological Physics, Weizmann Institute of Science, Rehovot 76100, Israel}
\author{Thomas M. Brenner}
\affiliation{Department of Chemical and Biological Physics, Weizmann Institute of Science, Rehovot 76100, Israel}
\author{Anna Kossoy}
\affiliation{Chemical Research Support, Weizmann Institute of Science, Rehovot 76100, Israel}
\author{Omer Yaffe}
\email{omer.yaffe@weizmann.ac.il}
\affiliation{Department of Chemical and Biological Physics, Weizmann Institute of Science, Rehovot 76100, Israel}
\author{Roman Korobko}
\email{roman.korobko@weizmann.ac.il}
\affiliation{Department of Chemical and Biological Physics, Weizmann Institute of Science, Rehovot 76100, Israel}

\date{\today}

\begin{abstract}

\dBO\ is one of the fastest known solid oxide ion conductors owing to its intrinsically defective fluorite-like structure with 25\% vacant sites in the O sub-lattice.
Numerous diffraction measurements and molecular dynamics simulations indicate that the Bi ions construct a cubic, fcc lattice, and the O ions are 
migrating through it. 
Nonetheless, in this study we present Raman scattering measurements that clearly show that the Bi sub-lattice preserves the monoclinic symmetry of the low temperature phase ($\alpha$-Bi$_2$O$_3$) up to the melting temperature of the crystal.  
The apparent contradiction between our observations and previous findings suggests that Bi ions oscillate between local minima of the effective potential energy surface.
These minima represent the monoclinic phase while the time-averaged structure is the cubic phase. 
We discuss the implication of these low-frequency oscillations on ion conduction.  

\end{abstract}
\maketitle

Fast solid oxide-ion conductors are attractive electrolyte materials for energy-related applications such as oxygen sensors, solid oxide fuel and electrolyzer cells\cite{Rickert1978, Mogensen2004,Goodenough2000, Famprikis2019}.
Amongst them, anion deficient fluorite structures have been found most suitable for efficient ion conduction.  Typical examples of fluorite oxide-ion conductors include stabilized cubic zirconia, doped ceria and the high-temperature ($>$ 727~$^{\circ}$C) phase of bismuth oxide (\dBO)\cite{Goodenough2000,Wachsman2011,Punn2006,KHARTON2004135}.
Bi$_{2}$O$_{3}$ has been extensively studied for decades\cite{WATANABE20052429,Drache2007,Yavo2016} due to its exceptionally high ionic conductivity in the $\delta$ phase (1~S\wn)\cite{Mamontov2016,Shuk1996,Sammes1999}.

\begin{figure}
\centering
\includegraphics[width=8.6cm, keepaspectratio=true]
{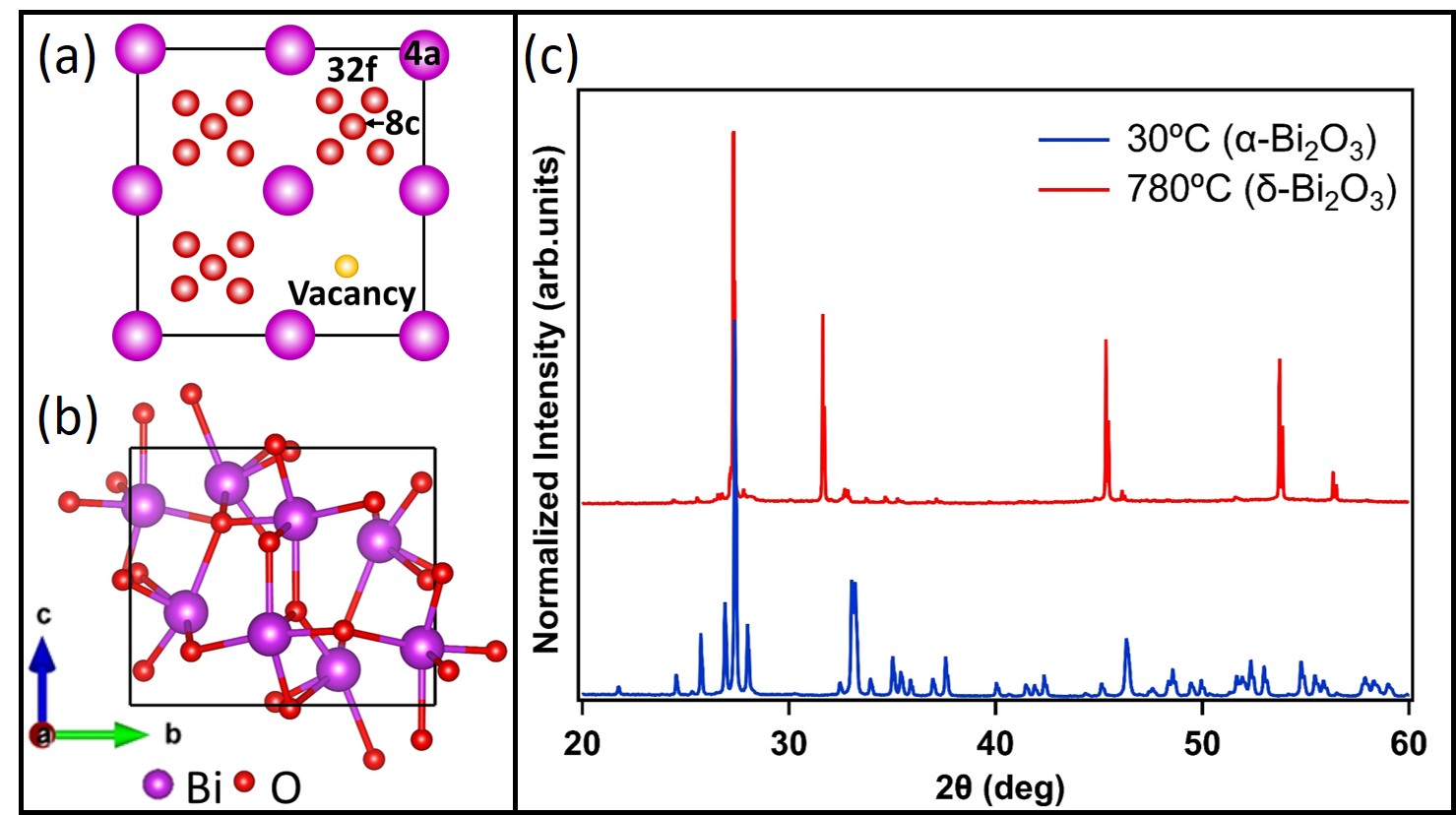}
\caption{(a) and (b) show the time-averaged crystal structures of the $\delta$ and $\alpha$-phase of \BO. (c) X-ray diffraction patterns of \BO\ for room temperature $\alpha$-phase (blue) and high temperature $\delta$-phase (red).}
\label{XRD}
\end{figure}

$\delta$--Bi$_{2}$O$_{3}$ exists at ambient pressure between 727 and 824~$^{\circ}$C. 
It has a defective fluorite structure with the time-average space group $Fm\bar{3}m$, with 25\% of intrinsic oxygen vacant sites.
Bi ions occupy the 4$a$ Wyckoff sites forming a face-centered cubic (fcc) sub-lattice (Fig.~\ref{XRD}(a).
The O ions partially occupy 32$f$ interstitial sites in addition to 8$c$ fluorite sites\cite{Yashima2003,MohnPRL,Shuk1996,doi:10.1021/ja065961d,wind_liquid-like_2017}. 
The stable phase at room temperature and atmospheric pressure is $\alpha$--Bi$_{2}$O$_{3}$, and has a monoclinic structure with the time-average space group $P2_{1}/c$ (Fig.~\ref{XRD}(b)).
Other, meta-stable polymorphs also exist (e.g., $\beta$, $\gamma$), depending on the cooling process the $\delta$-phase undergoes\cite{Shuk1996,Harwig1979121}. 
Due to its technological relevance, the $\alpha \rightarrow \delta$ phase transition was thoroughly investigated~\cite{YashimaNPD,19784440119}.
It is has been described as a diffusionless transition that involves the cooperative motion of the Bi ions from the monoclinic to the fluorite positions\cite{Yashima1996,19784440119}.

\begin{figure*}
\centering
\includegraphics[width=17cm, keepaspectratio=true]{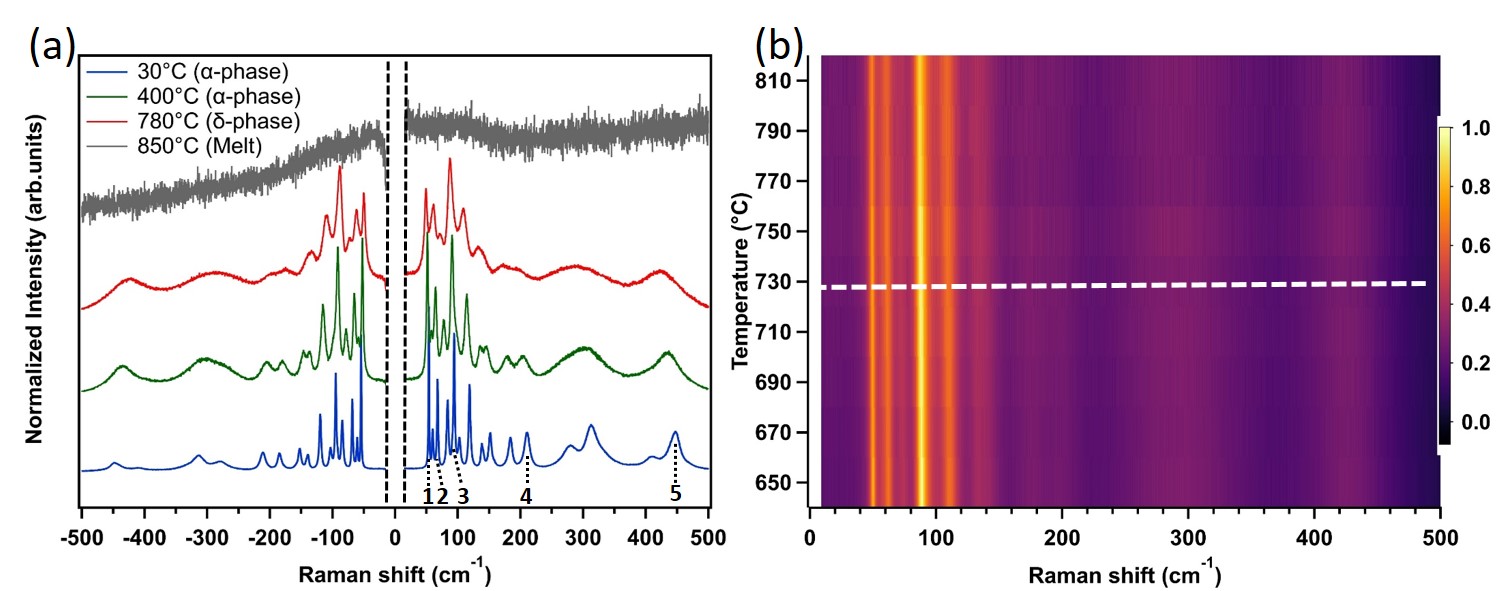}
\caption{(a) Raman spectra of \BO\ at temperatures corresponding to $\alpha$, $\delta$ and melted phases. The region between -10 and 10 \wn\ is cut off by notch filters. (b) Contour plot showing temperature dependence of Raman scattering (Stoke side) close to the phase transition ($T_{\alpha~\rightarrow~\delta}$ $\approx$ 727\degs C, shown by doted line). All spectra are normalized to their maximum intensity peak. No noticeable change in the spectra is observed across the phase transition temperature.}
\label{Raman}
\end{figure*}

In $\delta$--Bi$_{2}$O$_{3}$, the oxygen ions hop between oxygen vacant sites leading to its high ion conductivity.
The mechanism leading to the diffusion of the oxygen ions involves liquid-like motion between nearest-neighbor vacancies via 32$f$ sites~\cite{Mohn2009,Aidhy2008}.   
Importantly, neutron scattering studies have found that the Bi fcc sub-lattice is stable, where the Bi ions harmonically fluctuate around their equilibrium position~\cite{Aidhy2008,wind_liquid-like_2017,Goel2020}.
However, first principle calculations predict that a perfect fcc Bi sub-lattice will lead to metallic or semimetallic electronic structure~\cite{MohnPRL,Walsh2006}. 
This is at odds with experimentally measured band gap of 1.73~eV\cite{Fan_2006}.
Only computations of crystal structures where the Bi sub-lattice is of lower symmetry show a clear band gap~\cite{Matsumoto2010}.
This discrepancy indicates that the local environment of Bi ions in \dBO\ is still not well-understood.   

We therefore measure the temperature evolution of the Raman spectra across the Bi$_{2}$O$_{3}$ $\alpha \rightarrow \delta$ phase transition.
Our results show that this phase transition has no detectable effect on the Raman spectra as the monoclinic spectral features are maintained.
We resolve the discrepancy between our results and those which clearly observe this phase transition, by suggesting that the effective potential energy surface of the Bi ions has a double-well shape.
Finally, we discuss the implications of our findings on the interplay between ionic conductivity and anharmonic structural dynamics of the host lattice.

Fig.~\ref{XRD}(c) shows the powder XRD measurements at temperatures corresponding to $\alpha$ and $\delta$ phases. 
We observed a stable $\delta$-phase at 780 $^\circ$C.
These results are in agreement with previous reports\cite{Schroder2010,YashimaNPD,Zhu_2016,Klinkova2007}. 

Figure~\ref{Raman}(a) presents the Raman spectra measured at selected temperatures (measurements at 100 \degs C increments are shown in Fig.~S1\cite{SI} in the SI).
Our results are fully reversible with temperature. 
Notably, the spectra acquired above 500 \degs C were corrected for thermal radiation as discussed in the experimental part. 

Factor group analysis for the \aBO\ structure predicts 30 zone center Raman active modes (15A$_{g}$+15B$_{g}$)\cite{Pereira_2014,BETSCH1978505}.
We observe 18 Raman modes in our experiment.
The remaining modes in the spectra are not detected possibly due to 
the overlapping of broadened Raman peaks lying very close to each other or low Raman scattering cross section.

The room temperature Raman spectrum shows sharp peaks below 200~\wn\ and broad features above 200 \wn. Reported computations indicate that vibrational motion below 120 \wn\ is mostly related to the Bi ions while vibrations above 150 \wn\ correspond to the O ions~\cite{Denisov1997}.
The broadening of high frequency Raman modes was attributed to positional disorder 
and strong anharmonicity of the oxygen atoms~\cite{Pereira_2014,Denisov1997,BETSCH1978505}.

As temperature increases up to the phase transition temperature, we observe a gradual softening (i.e., shift to lower energies) and broadening of all observed peaks, which is attributed mainly to thermal expansion. 
We also observe a rising continuous background up to 500 \wn, even after subtraction of the thermal radiation.
This is most likely a second-order Raman signal, originating in inelastic scattering events including multiple phonons\cite{Cardona1982,Nimrod}.

The most striking observation in this study is that the Raman spectra show no abrupt change across the $\alpha\to\delta$ phase transition (Fig.~\ref{Raman}(b)), as expected in the case of an abrupt change in the crystal structure. Also, no triply degenerate T$_{2g}$ Raman mode, a characteristic of other fluorite-type solid oxide conductors, such as stabilized cubic zirconia and doped ceria, was observed ~\cite{schmitt_review_2020, FernandezLopez2001, Yashima1996}.

\begin{figure}
\centering
\includegraphics[width=8.6cm, keepaspectratio=true]
{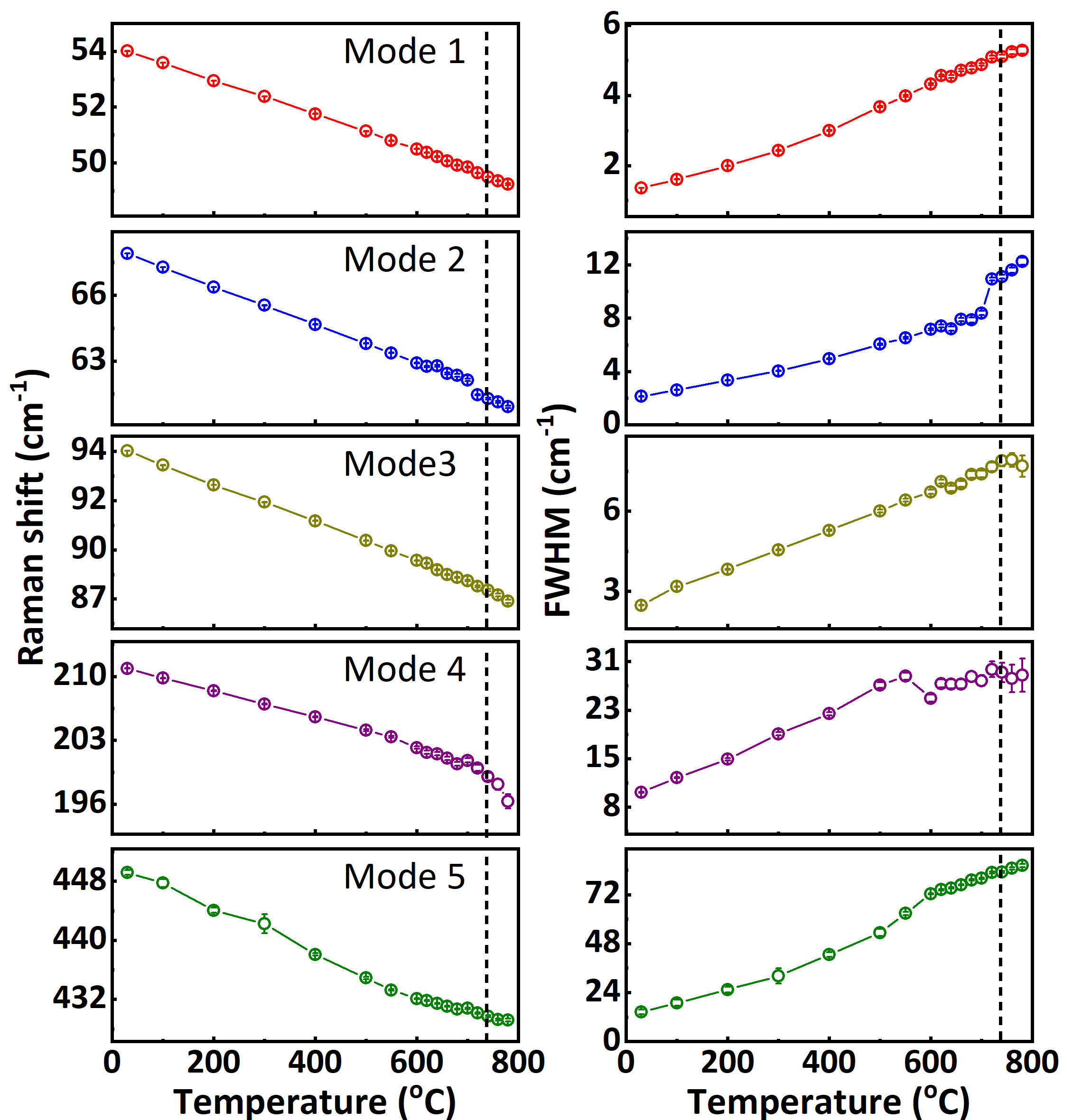}
\caption{\label{shift_broadening} Temperature dependence of mode positions (left) and FWHM (right) for prominent modes (1-5, indicated by dotted lines in Fig.~\ref{Raman}(a)) in \BO. The dashed lines represent the phase transition temperature.}
\end{figure}

To follow the temperature evolution of the Raman spectrum of \BO\ more accurately, we extract the temperature dependence of the vibrational frequency and full width at half maximum (FWHM) of each peak. These parameters were extracted by fitting the Raman spectra to the product of the Bose-Einstein distribution and a multi-damped Lorentz oscillator (see SI for details)\cite{SI}. Figure~\ref{shift_broadening} presents the results for the most prominent modes marked 1 to 5 on Fig.~\ref{Raman}a (the modes frequency are marked in the SI, Fig S3\cite{SI}). 
The results for the remaining modes are given in the SI, Figs. S5, S6 and S7\cite{SI}. Indeed within the error of the fit, we see no abrupt changes in the temperature dependence of the vibrational frequencies and FWHMs at the phase transition temperature and their linear trend 
for the $\alpha$-phase is maintained across the phase transition (see Figure~\ref{shift_broadening}). The abrupt changes that are observed, for instance in mode No. 2 at ~700 \degs C, are either due to an artifact of the fitting process or to the lower signal-to-noise ratio in higher temperatures (see SI for more details)\cite{SI}.
At 825~$^{\circ}$C the $\delta$--Bi$_{2}$O$_{3}$ melts.


These results indicate that the local environment of the Bi ions is unchanged across the phase transition.
However, diffraction experiments (including synchrotron, X-ray tube radiation and neutron)~\cite{Gattow,YashimaNPD,Harwig} and ionic conductivity measurements~\cite{Harwig1979121} clearly show that there is a sharp transition around 727~$^\circ$C.
This discrepancy can be resolved by considering an order-disorder phase transition of the Bi sub-lattice. 

\begin{figure}
\centering
\includegraphics[width=8cm, keepaspectratio=true]
{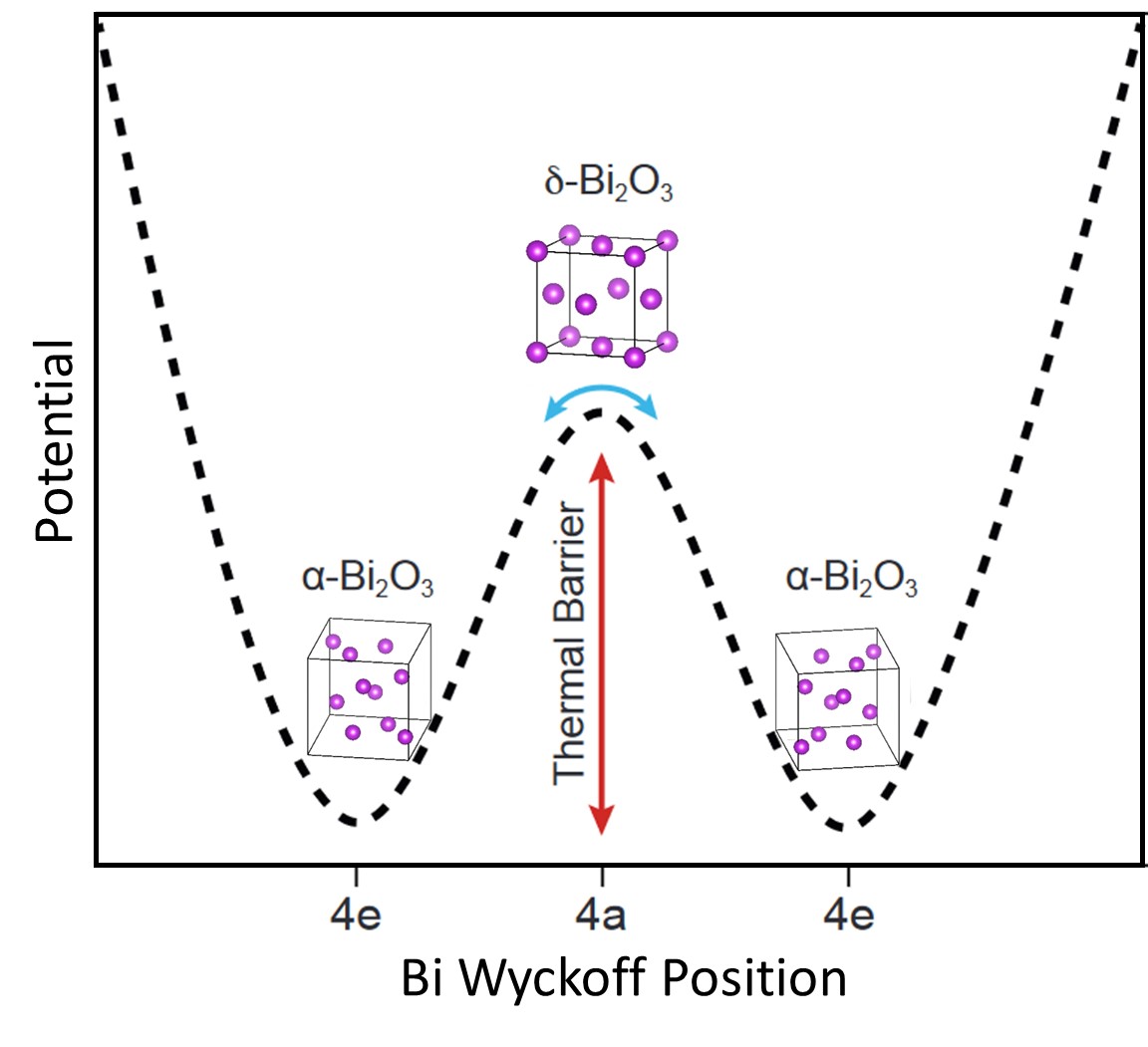}
\caption{\label{potential-well} Schematic representation of double-well potential model depicting the monoclinic and time averaged cubic configurations of the Bi sub-lattice.}
\end{figure}

Order-disorder transitions are a common mechanism of diffusionless phase transitions~\cite{Armstrong1989,Roleder2000,Xu2019,Scott1974,PhysRevLett.118.136001}. 
Such a transition can be described by a schematic double-well effective potential surface representing different, but equivalent, configurations of the Bi ion in a monoclinic structure (Fig.~\ref{potential-well}). 
When the thermal energy is low with respect to the energy barrier, the atoms oscillate locally within the well, where the effective harmonic potential holds. 
However, when the temperature is increased, larger thermal fluctuations increase the probability for atoms to hop between local minima (monoclinic structure), resulting in a change of the average crystal structure, but not in the local environment of the atoms. 
This hopping leads to strong dynamic disorder that occurs on time scales that are similar or longer than that of the vibrational motion~\cite{Scott1974,doi:10.1021/acsenergylett.6b00381,Picozzi_2008}. Our suggestion regarding dynamically stabilized cubic phase is supported by previously reported XAFS results, as they indicate that the $\delta$-Bi$_{2}$O$_{3}$ structure is not well-defined~\cite{Zhu_2016}.

Such dynamic disorder can have a strong effect on the Raman spectra for some materials\cite{Fontana_2020,Sharma,Blount1962,Brenner2020,Brenner2021,202107932,Menahem2021}. 
The fact that the Raman spectrum of $\delta$-Bi$_{2}$O$_{3}$ is essentially identical to that of $\alpha$-Bi$_{2}$O$_{3}$ indicates that the hopping of the Bi ions between wells of the effective potential surface occur a much longer timescale than that of the lattice vibrations ($\sim$THz).
Notably, the motion of the oxygen ions is more disordered than the Bi ions, as they can hop between lattice sites.
This is expressed by the relatively larger width (i.e., shorter vibrational lifetime) of the high-frequency modes which correspond to the O ions vibrations.


A real space description of this kind of order-disorder transition was already proposed in a structural meta-study by \citet{19784440119}.
Starting from the $\alpha$-phase Bi sub-lattice and its relation to a perfect fcc structure, the authors employed a ``least-motion" principle to determine the Bi shifts required for the $\alpha\to\delta$ transition.
While our current findings suggest no long-lived displacement takes place, it does however establish a possible relationship between the $\alpha$- and $\delta$-phase crystallographic orientations.
Specifically, the a*, b and c axes of \aBO\ constitute the cubic $\langle001\rangle$, $\langle\bar{1}10\rangle$ and $\langle110\rangle$ directions of \dBO, respectively.
Through coordinated motion of Bi atoms along the monoclinic c-axis, a perfect fcc projection on the fluorite $\left(101\right)$ plane is obtained.

We hypothesize that the dynamic collective motion of the Bi sub-lattice between the two monoclinic potential energy surface minima may facilitate the migration of O ions, as also suggested by~\citet{19784440119}.
This motion opens a conduction pathway for the large O ions which explains the sudden increase in ionic conductivity correlated with the XRD-detected $\alpha \rightarrow \delta$ phase transition.
Such a dynamic pathway in a double-well monoclinic configuration is more voluminous in comparison to the static cubic symmetry, where the cations are fixed (see Fig.~\ref{potential-well}). 
Also, our hypothesis provides explanation to the metallic or semimetallic electronic structure predicted by first principle calculations to the static cubic structure.
Since the \dBO\ oscillates between monoclinic structures and is cubic only on average, the ions are placed mostly at the monoclinic positions in space, so the electronic structure resembles more the monoclinic structure, as the static cubic picture does not exist.

Common mechanistic understanding of ion conduction in solids is understood primarily in terms of static host-structures (e.g., the Bi sub-lattice) and paths of ions through connected, open spaces in these crystals\cite{He2017,Goel2020,Physik}.
This standard view implicitly assumes only the mobile ions are in motion, and that the fluctuations of the atoms in the host lattice are negligible\cite{Mamontov2016,wind_liquid-like_2017,Goel2020}. 
We and others have recently suggested that the motion of the mobile O ions is strongly coupled to the structural dynamics of the rest of the lattice\cite{PhysRevMaterials.4.115402,Zhang2019,C7EE03364H,Krauskopf2018,Krauskopf2017,Kraft2017}. 
The anharmonic behavior of the Bi sub-lattice in \dBO\ is similar to the recently observed, anharmonic behavior of other solid ion conductors~\cite{Smith2020,Duchene2019,Adelstein2016,Kweon2017}. 

To conclude, we show that the crystals structure of \dBO\ is cubic only on average, as the Bi ions dynamically oscillate between monoclinic local minima on the effective double-well potential energy surface.
This explains the \dBO\ fluorite cubic structure yielded by XRD and its monoclinic Raman spectrum.
We suggest that the anharmonic dynamics of Bi cations are essential for the fast ionic conductivity, in addition to the high mobility of the O anions. 

\section*{{Experimental Methods}}
Bismuth oxide powder, purity of 99.999$\%$ by MaTecK GmbH, was used without any treatment.

XRD experiments were performed using Cu K-$\alpha$ radiation on SmartLab (Rigaku, Japan) diffractometer equipped with a rotating Cu anode, operating at 45 kV and 200 mA and with a HyPix-3000 two-dimensional detector in Bragg-Brentano geometry with variable incident slit. 
The beam was shaped by 2.5 deg solar before and after sample. Beam width was 5mm. The detector operated in 1D mode with scattering and receiving slits equal to 5 mm. XRD experiments were performed at ambient atmosphere. Bismuth oxide powder was heated in-situ with DHS 1100 heating stage by Anton Paar in ambient environment. Heating and cooling rate was 5 \degs C/min. The powder was kept at 790 \degs C for 1 hour before high temperature measurement. Diffraction patterns were measured between 2$\theta$ equal 10-120\degs,  with 0.5 \degs  C/min. Qualitative phase analysis was carried out using Jade Pro (MDI, USA) software. $\alpha$ and $\delta$ phases were matched with respective space groups in PDF-4+ (ICDD) database. 
 
Raman spectra were acquired using a custom built Raman system.
785~nm diode laser (TOPTICA Photonics AG) was used for excitation at a power of 8~mW, focused on the sample using a 10x objective (Nikon). 
Two volume holographic notch filters (each having OD$>$4
rejection with a spectral cut off $\pm$7 \wn~around 785~nm) discard the Rayleigh-scattered
light and allow measurements covering both Stokes and anti-Stokes sides of the spectrum.  
High-temperature measurements were carried out in a Linkam TS1000EV high temperature optical furnace under continuous flow of nitrogen.
The heating rate was kept constant at 5 \degs C/min from 30-700 \degs C, with an equilibration time of 10 minutes for each temperature. 
In the vicinity of the reported phase transition temperature (700-750 \degs C), the heating rate was reduced to 2 \degs C/min. 
Notably, high temperature (beyond 500\degs C) Raman spectra contained a uniform background due to thermal radiation (Fig. S2)\cite{SI}. 
To obtain the background corrected spectra at high temperatures, the thermal radiation was first measured without laser excitation in the same conditions before spectrum acquisition. 
Thereafter, the thermal radiation was subtracted from the Raman spectrum for the corresponding temperature.

\section*{Acknowledgements}
The authors would like to thank Dr. Lior Segev (WIS)
for invaluable software development. R.S. acknowledges FGS (WIS) for funding. O.Y. acknowledges
funding from ISF(1861/17), BSF (grant no. 2016650)
and ERC (850041-ANHARMONIC).

\clearpage
\newpage
\widetext
\begin{center}
	\large{\textbf{Dynamic disorder in Bi sub-lattice of \dBO}}
\end{center}

\begin{center}
	Rituraj Sharma$ ^{1} $, Nimrod Benshalom$ ^{1} $, Maor Asher$ ^{1} $, Thomas M. Brenner$ ^{1} $, Anna Kossi$ ^{2} $, Omer Yaffe$ ^{1*} $, Roman Korobko$ ^{1\ddagger} $ 
	\end{center}

\begin{center}
	\textit{$^1$Department of Chemical and Biological Physics, Weizmann Institute of Science, Rehovoth 76100, Israel\\
		$^2$Chemical Research Support, Weizmann Institute of Science, Rehovoth 76100, Israel}
\end{center}

\section*{SUPPLEMENTAL MATERIAL}

\setcounter{figure}{0}
\setcounter{page}{1}
\setcounter{equation}{0}
\setcounter{table}{0}
\setcounter{subsection}{0}

\renewcommand{\thefigure}{S\arabic{figure}}
\renewcommand{\thetable}{S\arabic{table}}
\renewcommand{\theequation}{S\arabic{equation}}
\renewcommand{\thesubsection}{\arabic{subsection}}
\renewcommand{\bibnumfmt}[1]{[S#1]}
\renewcommand{\citenumfont}[1]{S#1}

\noindent \subsection*{Fitting Procedure}

The Raman spectrum at each temperature was deconvolved to a product of the Bose-Einstein distribution ($ n_{BE} $) and multi-Lorentz line shape using the procedure used in Ref~\cite{Asher20p1908028,Sharma}, using a  customised code in Igor Pro 8. The experimentally measured Raman spectrum is expressed as a combination of Debye and a damped Lorentz oscillator term (eq.~\ref{s1}).

\begin{eqnarray}
	\label{s1}
	&I_{exp}(\nu,\nu_i,\Gamma_i)=c_{BE}(\nu)\left(\frac{c_0\left| \nu\right|\Gamma_0}{\nu^2+\Gamma_0^2} + \sum_{i=1}^{n} c_i\frac{\left| \nu\right| \left| \nu_i\right| \Gamma_i^2}{\nu^2\Gamma_i^2 + (\nu^2 - \nu_i^2)^2}\right),
\end{eqnarray}
\\	
\noindent where $ \nu $ is the spectral shift, $ \nu_i $s are the resonance energies of the Lorentz oscillators, $\Gamma_0$ and  $\Gamma_i$ are the damping coefficients of the Debye relaxation and Lorentz oscillators respectively, $ c_{0} $ and $ c_{i} $s are unitless fitting parameters for the intensities of the Debye and Lorentz oscillator components respectively. The spectral shift $ \nu $, the parameters  $ \nu_i $ and $\Gamma_i $ are in wavenumber units. $ c_{BE}(\nu) $ includes the Bose-Einstein distribution of thermal population where 
\begin{equation*}
	c_{BE}(\nu) = \begin{cases}
		n_{BE} +1 & $for Stokes scattering,$ \\
		n_{BE} & $for Anti-Stokes scattering.$
	\end{cases}
\end{equation*}
with \textit{n} being the Bose-Einstein distribution.
The spectral artifacts due to the notch filter around 0 cm$^{-1}$ were omitted and only the Stokes scattering was fitted to find the positions and widths of the peaks.

\begin{figure}[h]
	\centering
	\includegraphics[width=8.5cm, keepaspectratio=true]
	{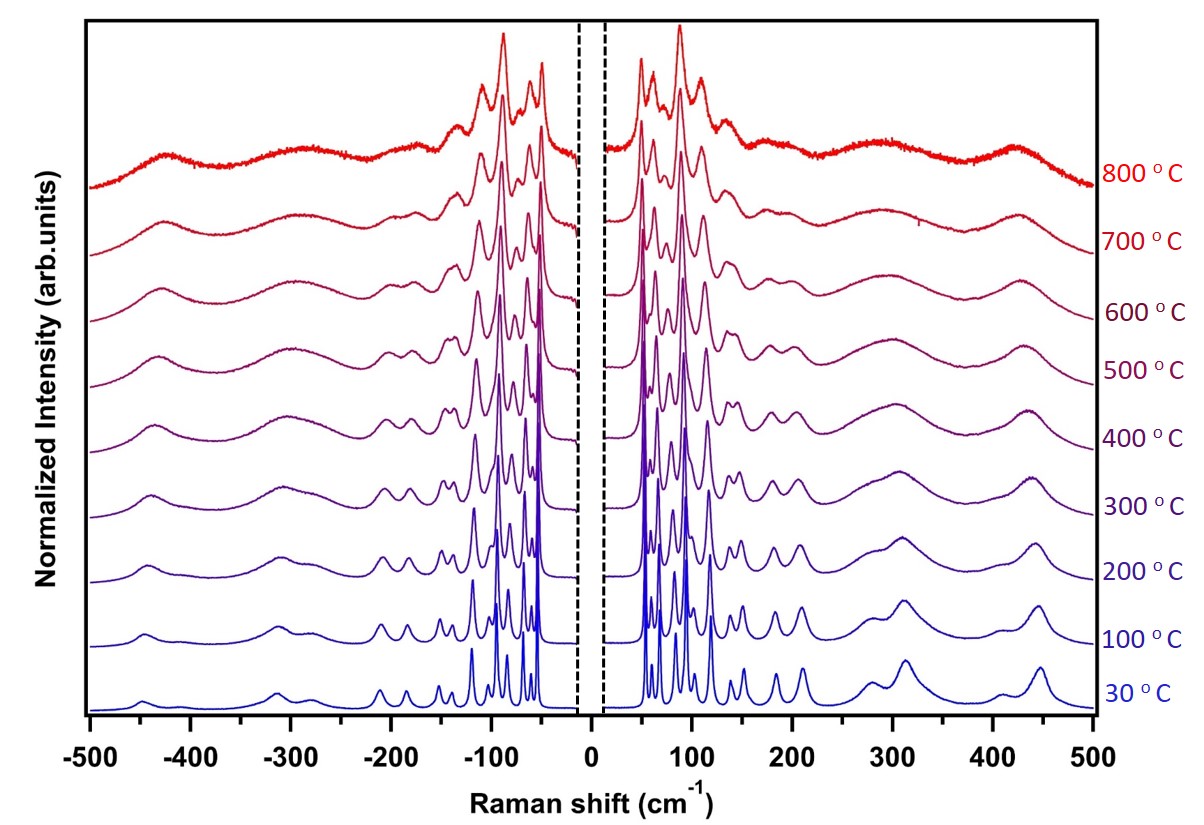}
	\caption{Temperature dependence of Raman spectra in \BO. The dotted line in the right figure is the reported phase transition temperature (727\degs C).}
	\label{}
\end{figure}

\begin{figure}
	\centering
	\includegraphics[width=15cm, keepaspectratio=true]
	{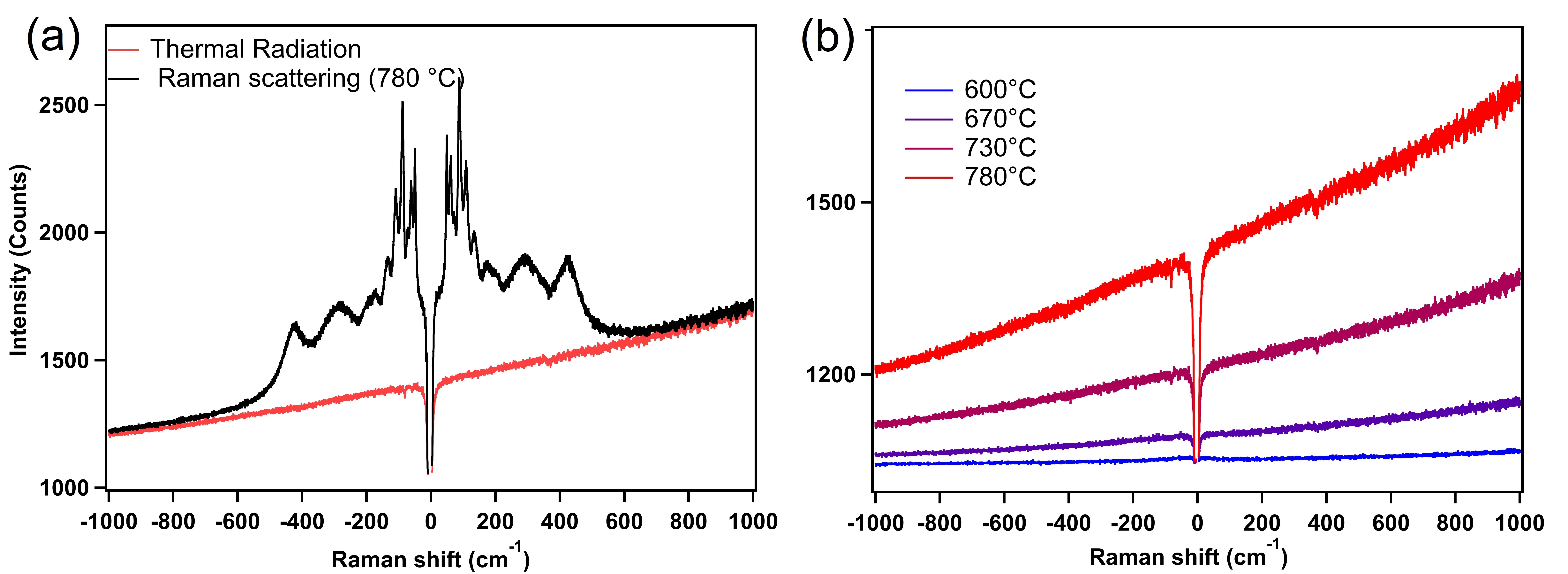}
	\caption{(a) Thermal radiation background (black) captured without excitation and its effect on the Raman spectrum at 780\degs C (red). The thermal radiation is subtracted from the Raman spectrum to get the base corrected spectrum. (b) Thermal radiation background captured without excitation at different temperatures.}
	\label{}
\end{figure}

\begin{figure}
	\centering
	\includegraphics[width=8.6cm, keepaspectratio=true]
	{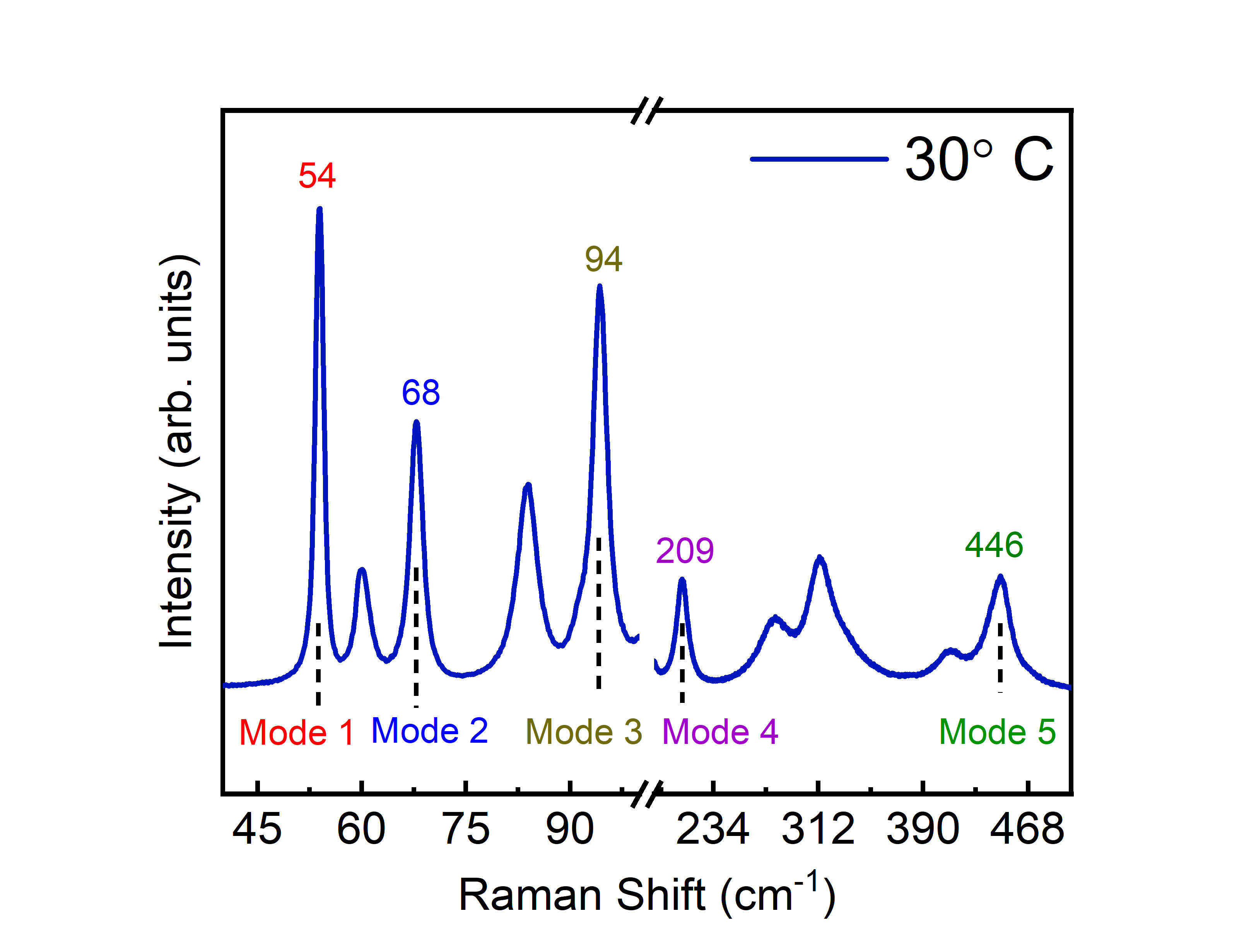}
	\caption{Frequencies of modes used in Fig. 3 of main text}
	\label{}
\end{figure}

\begin{figure}
	\centering
	\includegraphics[width=8.6cm, keepaspectratio=true]
	{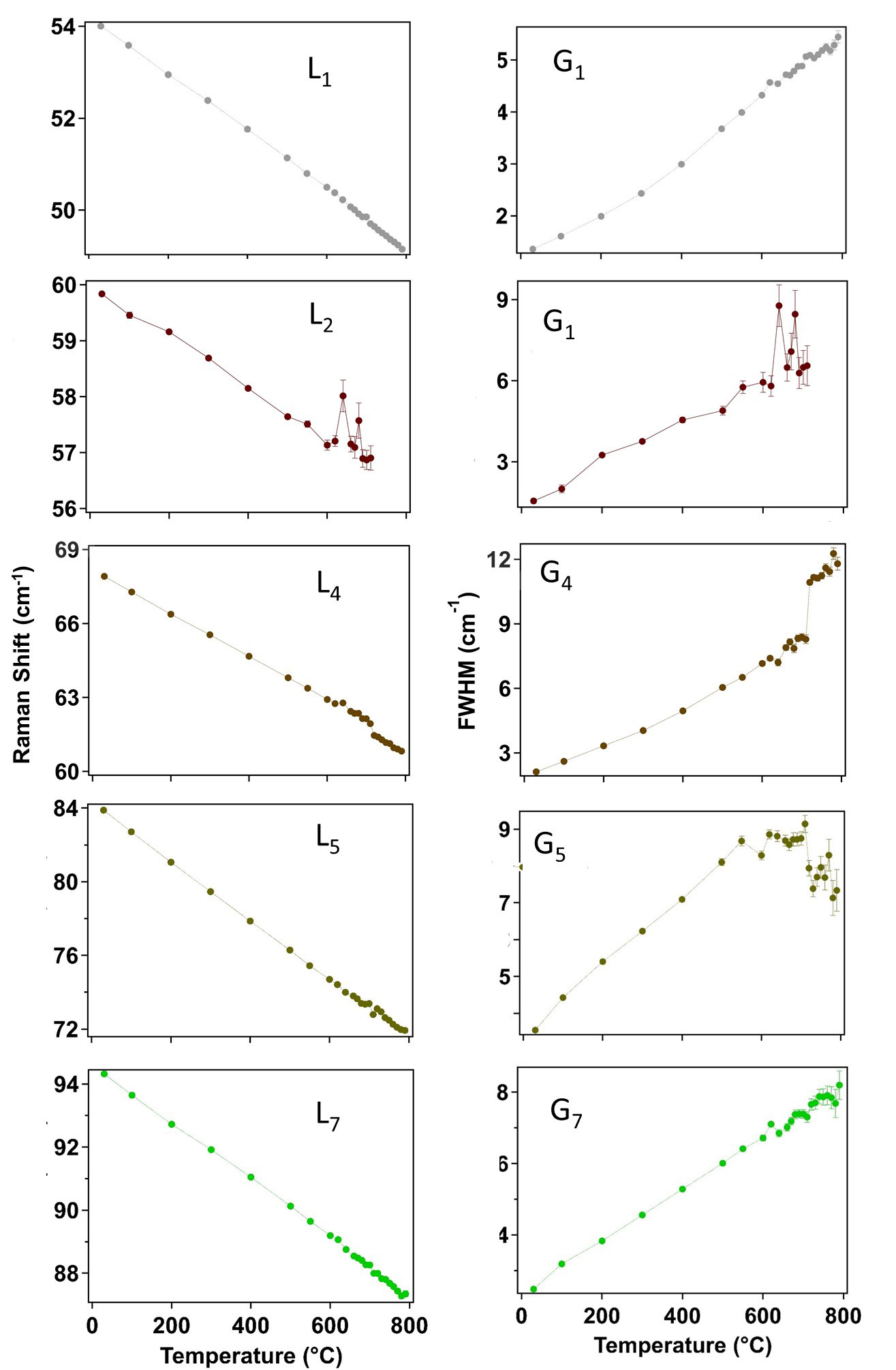}
	\caption{Temperature evolution of peak shift  (right) and FWHM (left) in \BO}
	\label{}
\end{figure}

\begin{figure}
	\centering
	\includegraphics[width=8.6cm, keepaspectratio=true]
	{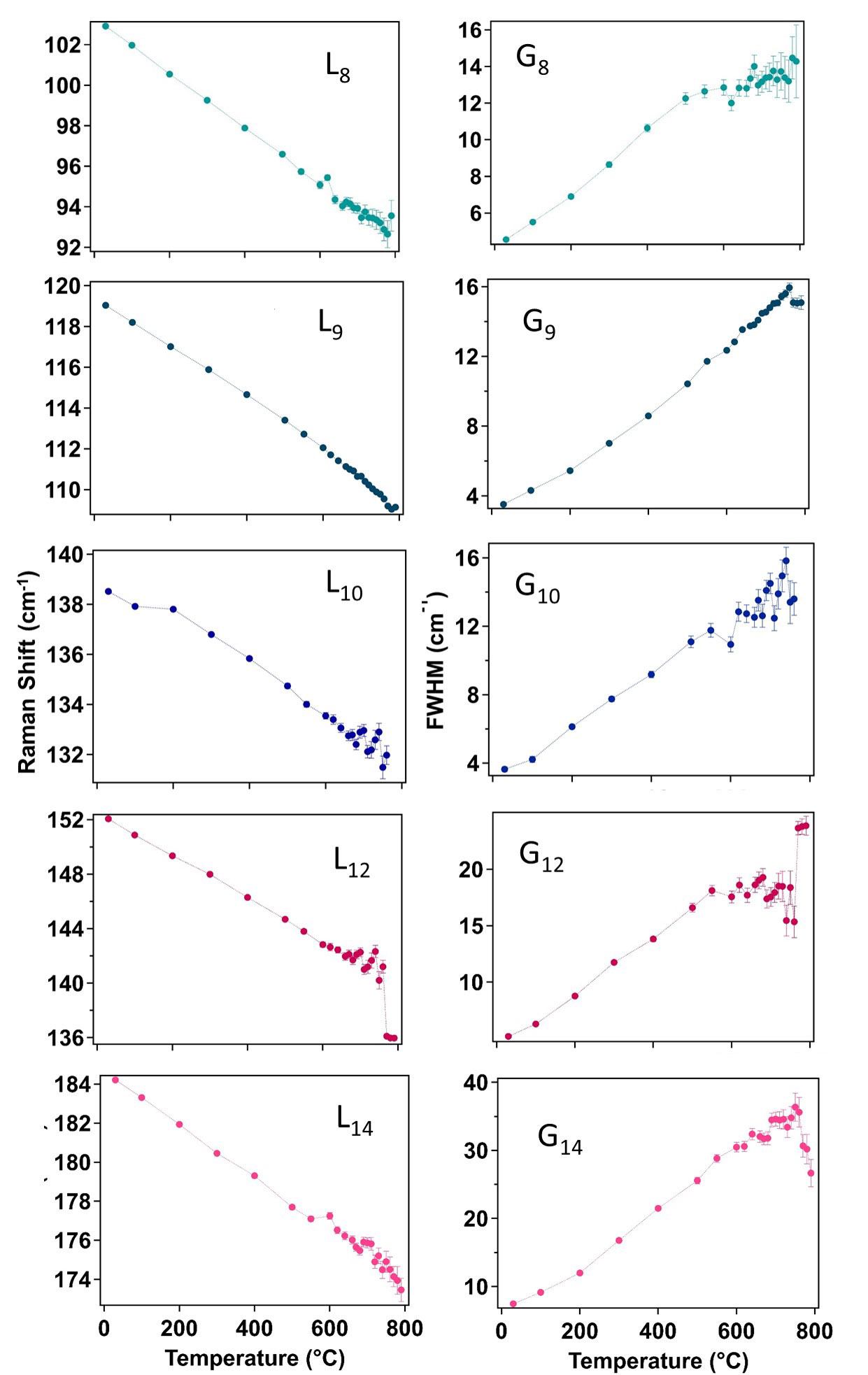}
	\caption{Temperature evolution of peak shift  (right) and FWHM (left) in \BO}
	\label{}
\end{figure}

\begin{figure}
	\centering
	\includegraphics[width=8.6cm, keepaspectratio=true]
	{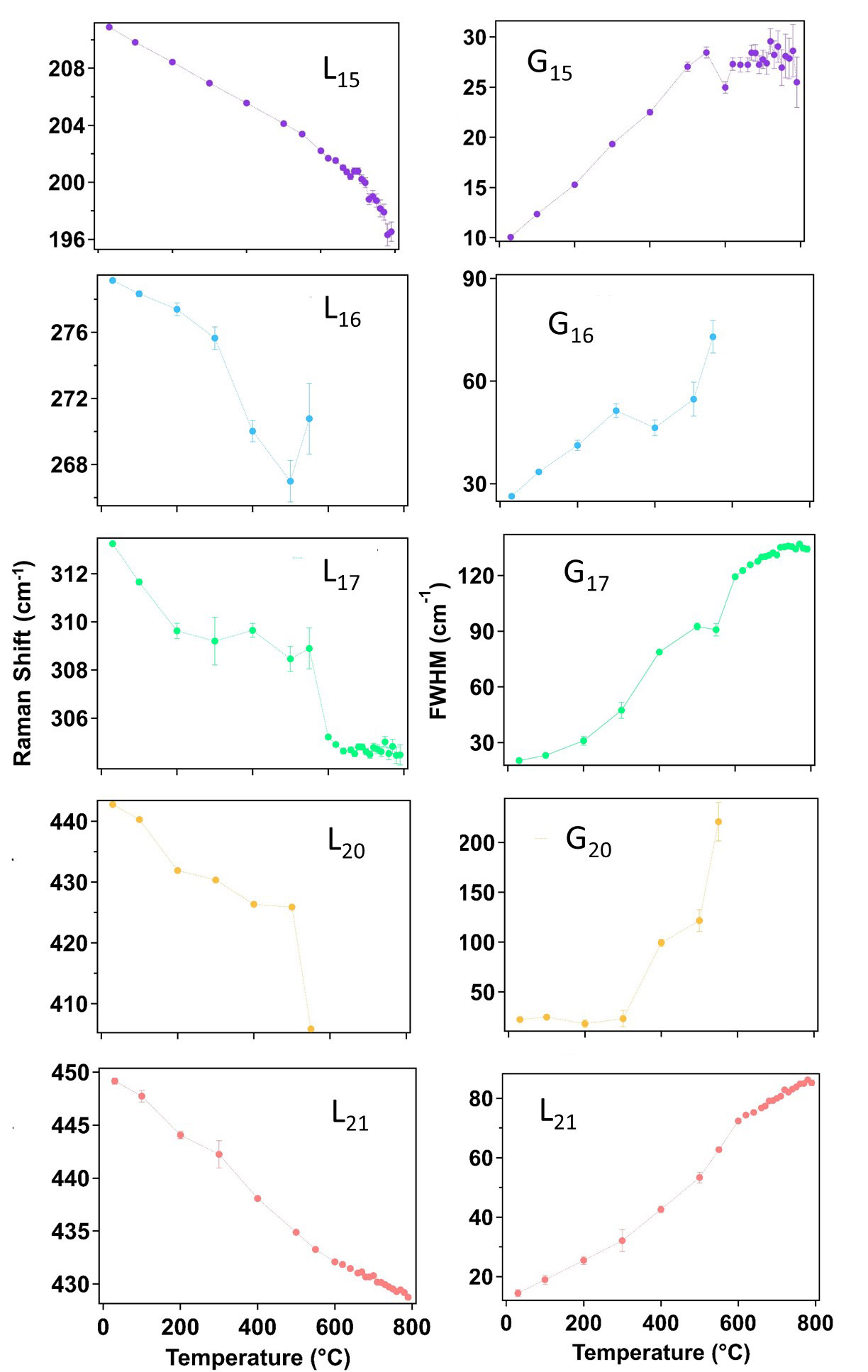}
	\caption{Temperature evolution of peak shift  (right) and FWHM (left) in \BO}
	\label{}
\end{figure}
\newpage

%

\begin{thebibliography}{66}%
	\makeatletter
	\providecommand \@ifxundefined [1]{%
		\@ifx{#1\undefined}
	}%
	\providecommand \@ifnum [1]{%
		\ifnum #1\expandafter \@firstoftwo
		\else \expandafter \@secondoftwo
		\fi
	}%
	\providecommand \@ifx [1]{%
		\ifx #1\expandafter \@firstoftwo
		\else \expandafter \@secondoftwo
		\fi
	}%
	\providecommand \natexlab [1]{#1}%
	\providecommand \enquote  [1]{``#1''}%
	\providecommand \bibnamefont  [1]{#1}%
	\providecommand \bibfnamefont [1]{#1}%
	\providecommand \citenamefont [1]{#1}%
	\providecommand \href@noop [0]{\@secondoftwo}%
	\providecommand \href [0]{\begingroup \@sanitize@url \@href}%
	\providecommand \@href[1]{\@@startlink{#1}\@@href}%
	\providecommand \@@href[1]{\endgroup#1\@@endlink}%
	\providecommand \@sanitize@url [0]{\catcode `\\12\catcode `\$12\catcode
		`\&12\catcode `\#12\catcode `\^12\catcode `\_12\catcode `\%12\relax}%
	\providecommand \@@startlink[1]{}%
	\providecommand \@@endlink[0]{}%
	\providecommand \url  [0]{\begingroup\@sanitize@url \@url }%
	\providecommand \@url [1]{\endgroup\@href {#1}{\urlprefix }}%
	\providecommand \urlprefix  [0]{URL }%
	\providecommand \Eprint [0]{\href }%
	\providecommand \doibase [0]{https://doi.org/}%
	\providecommand \selectlanguage [0]{\@gobble}%
	\providecommand \bibinfo  [0]{\@secondoftwo}%
	\providecommand \bibfield  [0]{\@secondoftwo}%
	\providecommand \translation [1]{[#1]}%
	\providecommand \BibitemOpen [0]{}%
	\providecommand \bibitemStop [0]{}%
	\providecommand \bibitemNoStop [0]{.\EOS\space}%
	\providecommand \EOS [0]{\spacefactor3000\relax}%
	\providecommand \BibitemShut  [1]{\csname bibitem#1\endcsname}%
	\let\auto@bib@innerbib\@empty
	\bibitem [{\citenamefont {Rickert}(1978)}]{Rickert1978}%
	\BibitemOpen
	\bibfield  {author} {\bibinfo {author} {\bibfnamefont {H.}~\bibnamefont
			{Rickert}},\ }\bibfield  {title} {\bibinfo {title} {{Solid Ionic Conductors:
				Principles and Applications}},\ }\href
	{https://doi.org/10.1002/anie.197800371} {\bibfield  {journal} {\bibinfo
			{journal} {Angew. Chem. Int. Ed.}\ }\textbf {\bibinfo {volume} {17}},\
		\bibinfo {pages} {37} (\bibinfo {year} {1978})}\BibitemShut {NoStop}%
	\bibitem [{\citenamefont {Mogensen}\ \emph {et~al.}(2004)\citenamefont
		{Mogensen}, \citenamefont {Lybye}, \citenamefont {Bonanos}, \citenamefont
		{Hendriksen},\ and\ \citenamefont {Poulsen}}]{Mogensen2004}%
	\BibitemOpen
	\bibfield  {author} {\bibinfo {author} {\bibfnamefont {M.}~\bibnamefont
			{Mogensen}}, \bibinfo {author} {\bibfnamefont {D.}~\bibnamefont {Lybye}},
		\bibinfo {author} {\bibfnamefont {N.}~\bibnamefont {Bonanos}}, \bibinfo
		{author} {\bibfnamefont {P.~V.}\ \bibnamefont {Hendriksen}},\ and\ \bibinfo
		{author} {\bibfnamefont {F.~W.}\ \bibnamefont {Poulsen}},\ }\bibfield
	{title} {\bibinfo {title} {{Factors controlling the oxide ion conductivity of
				fluorite and perovskite structured oxides}},\ }\href
	{https://doi.org/10.1016/j.ssi.2004.07.036} {\bibfield  {journal} {\bibinfo
			{journal} {Solid State Ionics}\ }\textbf {\bibinfo {volume} {174}},\ \bibinfo
		{pages} {279} (\bibinfo {year} {2004})}\BibitemShut {NoStop}%
	\bibitem [{\citenamefont {Goodenough}(2000)}]{Goodenough2000}%
	\BibitemOpen
	\bibfield  {author} {\bibinfo {author} {\bibfnamefont {J.~B.}\ \bibnamefont
			{Goodenough}},\ }\bibfield  {title} {\bibinfo {title} {{Oxide-ion conductors
				by design}},\ }\href {https://doi.org/10.1038/35009177} {\bibfield  {journal}
		{\bibinfo  {journal} {Nature}\ }\textbf {\bibinfo {volume} {404}},\ \bibinfo
		{pages} {821} (\bibinfo {year} {2000})}\BibitemShut {NoStop}%
	\bibitem [{\citenamefont {Famprikis}\ \emph {et~al.}(2019)\citenamefont
		{Famprikis}, \citenamefont {Canepa}, \citenamefont {Dawson}, \citenamefont
		{Islam},\ and\ \citenamefont {Masquelier}}]{Famprikis2019}%
	\BibitemOpen
	\bibfield  {author} {\bibinfo {author} {\bibfnamefont {T.}~\bibnamefont
			{Famprikis}}, \bibinfo {author} {\bibfnamefont {P.}~\bibnamefont {Canepa}},
		\bibinfo {author} {\bibfnamefont {J.~A.}\ \bibnamefont {Dawson}}, \bibinfo
		{author} {\bibfnamefont {M.~S.}\ \bibnamefont {Islam}},\ and\ \bibinfo
		{author} {\bibfnamefont {C.}~\bibnamefont {Masquelier}},\ }\bibfield  {title}
	{\bibinfo {title} {{Fundamentals of inorganic solid-state electrolytes for
				batteries}},\ }\href {https://doi.org/10.1038/s41563-019-0431-3} {\bibfield
		{journal} {\bibinfo  {journal} {Nat. Mater.}\ }\textbf {\bibinfo {volume}
			{18}},\ \bibinfo {pages} {1278} (\bibinfo {year} {2019})}\BibitemShut
	{NoStop}%
	\bibitem [{\citenamefont {Wachsman}\ and\ \citenamefont
		{Lee}(2011)}]{Wachsman2011}%
	\BibitemOpen
	\bibfield  {author} {\bibinfo {author} {\bibfnamefont {E.~D.}\ \bibnamefont
			{Wachsman}}\ and\ \bibinfo {author} {\bibfnamefont {K.~T.}\ \bibnamefont
			{Lee}},\ }\bibfield  {title} {\bibinfo {title} {{Lowering the temperature of
				solid oxide fuel cells}},\ }\href {https://doi.org/10.1126/science.1204090}
	{\bibfield  {journal} {\bibinfo  {journal} {Science}\ }\textbf {\bibinfo
			{volume} {334}},\ \bibinfo {pages} {935} (\bibinfo {year}
		{2011})}\BibitemShut {NoStop}%
	\bibitem [{\citenamefont {Punn}\ \emph
		{et~al.}(2006{\natexlab{a}})\citenamefont {Punn}, \citenamefont {Feteira},
		\citenamefont {Sinclair},\ and\ \citenamefont {Greaves}}]{Punn2006}%
	\BibitemOpen
	\bibfield  {author} {\bibinfo {author} {\bibfnamefont {R.}~\bibnamefont
			{Punn}}, \bibinfo {author} {\bibfnamefont {A.~M.}\ \bibnamefont {Feteira}},
		\bibinfo {author} {\bibfnamefont {D.~C.}\ \bibnamefont {Sinclair}},\ and\
		\bibinfo {author} {\bibfnamefont {C.}~\bibnamefont {Greaves}},\ }\bibfield
	{title} {\bibinfo {title} {{Enhanced Oxide Ion Conductivity in Stabilized
				$\delta$-Bi2O3}},\ }\href {https://doi.org/10.1021/ja065961d} {\bibfield
		{journal} {\bibinfo  {journal} {J. Am. Chem. Soc.}\ }\textbf {\bibinfo
			{volume} {128}},\ \bibinfo {pages} {15386} (\bibinfo {year}
		{2006}{\natexlab{a}})}\BibitemShut {NoStop}%
	\bibitem [{\citenamefont {Kharton}\ \emph {et~al.}(2004)\citenamefont
		{Kharton}, \citenamefont {Marques},\ and\ \citenamefont
		{Atkinson}}]{KHARTON2004135}%
	\BibitemOpen
	\bibfield  {author} {\bibinfo {author} {\bibfnamefont {V.~V.}\ \bibnamefont
			{Kharton}}, \bibinfo {author} {\bibfnamefont {F.~M.~B.}\ \bibnamefont
			{Marques}},\ and\ \bibinfo {author} {\bibfnamefont {A.}~\bibnamefont
			{Atkinson}},\ }\bibfield  {title} {\bibinfo {title} {{Transport properties of
				solid oxide electrolyte ceramics: a brief review}},\ }\href
	{https://doi.org/https://doi.org/10.1016/j.ssi.2004.06.015} {\bibfield
		{journal} {\bibinfo  {journal} {Solid State Ion.}\ }\textbf {\bibinfo
			{volume} {174}},\ \bibinfo {pages} {135} (\bibinfo {year}
		{2004})}\BibitemShut {NoStop}%
	\bibitem [{\citenamefont {Watanabe}\ and\ \citenamefont
		{Sekita}(2005)}]{WATANABE20052429}%
	\BibitemOpen
	\bibfield  {author} {\bibinfo {author} {\bibfnamefont {A.}~\bibnamefont
			{Watanabe}}\ and\ \bibinfo {author} {\bibfnamefont {M.}~\bibnamefont
			{Sekita}},\ }\bibfield  {title} {\bibinfo {title} {Stabilized $ \delta$-bi2o3
			phase in the system bi2o3–er2o3–wo3 and its oxide-ion conduction},\
	}\href {https://doi.org/https://doi.org/10.1016/j.ssi.2005.02.027} {\bibfield
		{journal} {\bibinfo  {journal} {Solid State Ionics}\ }\textbf {\bibinfo
			{volume} {176}},\ \bibinfo {pages} {2429} (\bibinfo {year} {2005})},\
	\bibinfo {note} {30th Symposium on Solid State Ionics in Japan}\BibitemShut
	{NoStop}%
	\bibitem [{\citenamefont {Drache}\ \emph {et~al.}(2007)\citenamefont {Drache},
		\citenamefont {Roussel},\ and\ \citenamefont {Wignacourt}}]{Drache2007}%
	\BibitemOpen
	\bibfield  {author} {\bibinfo {author} {\bibfnamefont {M.}~\bibnamefont
			{Drache}}, \bibinfo {author} {\bibfnamefont {P.}~\bibnamefont {Roussel}},\
		and\ \bibinfo {author} {\bibfnamefont {J.-P.}\ \bibnamefont {Wignacourt}},\
	}\href {https://doi.org/10.1021/cr050977s} {\bibfield  {journal} {\bibinfo
			{journal} {Chem. Rev.}\ }\textbf {\bibinfo {volume} {107}},\ \bibinfo {pages}
		{80} (\bibinfo {year} {2007})}\BibitemShut {NoStop}%
	\bibitem [{\citenamefont {Yavo}\ \emph {et~al.}(2016)\citenamefont {Yavo},
		\citenamefont {Smith}, \citenamefont {Yeheskel}, \citenamefont {Cohen},
		\citenamefont {Korobko}, \citenamefont {Wachtel}, \citenamefont {Slater},\
		and\ \citenamefont {Lubomirsky}}]{Yavo2016}%
	\BibitemOpen
	\bibfield  {author} {\bibinfo {author} {\bibfnamefont {N.}~\bibnamefont
			{Yavo}}, \bibinfo {author} {\bibfnamefont {A.~D.}\ \bibnamefont {Smith}},
		\bibinfo {author} {\bibfnamefont {O.}~\bibnamefont {Yeheskel}}, \bibinfo
		{author} {\bibfnamefont {S.}~\bibnamefont {Cohen}}, \bibinfo {author}
		{\bibfnamefont {R.}~\bibnamefont {Korobko}}, \bibinfo {author} {\bibfnamefont
			{E.}~\bibnamefont {Wachtel}}, \bibinfo {author} {\bibfnamefont {P.~R.}\
			\bibnamefont {Slater}},\ and\ \bibinfo {author} {\bibfnamefont
			{I.}~\bibnamefont {Lubomirsky}},\ }\bibfield  {title} {\bibinfo {title}
		{{Large Nonclassical Electrostriction in (Y, Nb)-Stabilized
				$\delta$-Bi2O3}},\ }\href {https://doi.org/10.1002/adfm.201503942} {\bibfield
		{journal} {\bibinfo  {journal} {Adv. Funct. Mater}\ }\textbf {\bibinfo
			{volume} {26}},\ \bibinfo {pages} {1138} (\bibinfo {year}
		{2016})}\BibitemShut {NoStop}%
	\bibitem [{\citenamefont {Mamontov}(2016)}]{Mamontov2016}%
	\BibitemOpen
	\bibfield  {author} {\bibinfo {author} {\bibfnamefont {E.}~\bibnamefont
			{Mamontov}},\ }\bibfield  {title} {\bibinfo {title} {{Fast oxygen diffusion
				in bismuth oxide probed by quasielastic neutron scattering}},\ }\href
	{https://doi.org/10.1016/j.ssi.2016.09.022} {\bibfield  {journal} {\bibinfo
			{journal} {Solid State Ionics}\ }\textbf {\bibinfo {volume} {296}},\ \bibinfo
		{pages} {158} (\bibinfo {year} {2016})}\BibitemShut {NoStop}%
	\bibitem [{\citenamefont {Shuk}\ \emph {et~al.}(1996)\citenamefont {Shuk},
		\citenamefont {Wiemh{\"{o}}fer}, \citenamefont {Guth}, \citenamefont
		{G{\"{o}}pel},\ and\ \citenamefont {Greenblatt}}]{Shuk1996}%
	\BibitemOpen
	\bibfield  {author} {\bibinfo {author} {\bibfnamefont {P.}~\bibnamefont
			{Shuk}}, \bibinfo {author} {\bibfnamefont {H.~D.}\ \bibnamefont
			{Wiemh{\"{o}}fer}}, \bibinfo {author} {\bibfnamefont {U.}~\bibnamefont
			{Guth}}, \bibinfo {author} {\bibfnamefont {W.}~\bibnamefont {G{\"{o}}pel}},\
		and\ \bibinfo {author} {\bibfnamefont {M.}~\bibnamefont {Greenblatt}},\
	}\bibfield  {title} {\bibinfo {title} {{Oxide ion conducting solid
				electrolytes based on Bi2O3}},\ }\href
	{https://doi.org/10.1016/0167-2738(96)00348-7} {\bibfield  {journal}
		{\bibinfo  {journal} {Solid State Ionics}\ }\textbf {\bibinfo {volume}
			{89}},\ \bibinfo {pages} {179} (\bibinfo {year} {1996})}\BibitemShut
	{NoStop}%
	\bibitem [{\citenamefont {Sammes}\ \emph {et~al.}(1999)\citenamefont {Sammes},
		\citenamefont {Tompsett}, \citenamefont {N{\"{a}}fe},\ and\ \citenamefont
		{Aldinger}}]{Sammes1999}%
	\BibitemOpen
	\bibfield  {author} {\bibinfo {author} {\bibfnamefont {N.~M.}\ \bibnamefont
			{Sammes}}, \bibinfo {author} {\bibfnamefont {G.~A.}\ \bibnamefont
			{Tompsett}}, \bibinfo {author} {\bibfnamefont {H.}~\bibnamefont
			{N{\"{a}}fe}},\ and\ \bibinfo {author} {\bibfnamefont {F.}~\bibnamefont
			{Aldinger}},\ }\bibfield  {title} {\bibinfo {title} {{Bismuth based oxide
				electrolytes - Structure and ionic conductivity}},\ }\href
	{https://doi.org/10.1016/S0955-2219(99)00009-6} {\bibfield  {journal}
		{\bibinfo  {journal} {J. Eur. Ceram. Soc}\ }\textbf {\bibinfo {volume}
			{19}},\ \bibinfo {pages} {1801} (\bibinfo {year} {1999})}\BibitemShut
	{NoStop}%
	\bibitem [{\citenamefont {Yashima}\ and\ \citenamefont
		{Ishimura}(2003)}]{Yashima2003}%
	\BibitemOpen
	\bibfield  {author} {\bibinfo {author} {\bibfnamefont {M.}~\bibnamefont
			{Yashima}}\ and\ \bibinfo {author} {\bibfnamefont {D.}~\bibnamefont
			{Ishimura}},\ }\bibfield  {title} {\bibinfo {title} {{Crystal structure and
				disorder of the fast oxide-ion conductor cubic Bi2O3}},\ }\href
	{https://doi.org/10.1016/j.cplett.2003.07.014} {\bibfield  {journal}
		{\bibinfo  {journal} {Chem. Phys. Lett.}\ }\textbf {\bibinfo {volume}
			{378}},\ \bibinfo {pages} {395} (\bibinfo {year} {2003})}\BibitemShut
	{NoStop}%
	\bibitem [{\citenamefont {Mohn}\ \emph
		{et~al.}(2009{\natexlab{a}})\citenamefont {Mohn}, \citenamefont {St\o{}len},
		\citenamefont {Norberg},\ and\ \citenamefont {Hull}}]{MohnPRL}%
	\BibitemOpen
	\bibfield  {author} {\bibinfo {author} {\bibfnamefont {C.~E.}\ \bibnamefont
			{Mohn}}, \bibinfo {author} {\bibfnamefont {S.}~\bibnamefont {St\o{}len}},
		\bibinfo {author} {\bibfnamefont {S.~T.}\ \bibnamefont {Norberg}},\ and\
		\bibinfo {author} {\bibfnamefont {S.}~\bibnamefont {Hull}},\ }\href
	{https://doi.org/10.1103/PhysRevLett.102.155502} {\bibfield  {journal}
		{\bibinfo  {journal} {Phys. Rev. Lett.}\ }\textbf {\bibinfo {volume} {102}},\
		\bibinfo {pages} {155502} (\bibinfo {year} {2009}{\natexlab{a}})}\BibitemShut
	{NoStop}%
	\bibitem [{\citenamefont {Punn}\ \emph
		{et~al.}(2006{\natexlab{b}})\citenamefont {Punn}, \citenamefont {Feteira},
		\citenamefont {Sinclair},\ and\ \citenamefont
		{Greaves}}]{doi:10.1021/ja065961d}%
	\BibitemOpen
	\bibfield  {author} {\bibinfo {author} {\bibfnamefont {R.}~\bibnamefont
			{Punn}}, \bibinfo {author} {\bibfnamefont {A.~M.}\ \bibnamefont {Feteira}},
		\bibinfo {author} {\bibfnamefont {D.~C.}\ \bibnamefont {Sinclair}},\ and\
		\bibinfo {author} {\bibfnamefont {C.}~\bibnamefont {Greaves}},\ }\bibfield
	{title} {\bibinfo {title} {Enhanced oxide ion conductivity in stabilized
			bi2o3},\ }\href {https://doi.org/10.1021/ja065961d} {\bibfield  {journal}
		{\bibinfo  {journal} {J. Am. Chem. Soc.}\ }\textbf {\bibinfo {volume}
			{128}},\ \bibinfo {pages} {15386} (\bibinfo {year} {2006}{\natexlab{b}})},\
	\bibinfo {note} {pMID: 17132000}\BibitemShut {NoStop}%
	\bibitem [{\citenamefont {Wind}\ \emph {et~al.}(2017)\citenamefont {Wind},
		\citenamefont {Mole}, \citenamefont {Yu},\ and\ \citenamefont
		{Ling}}]{wind_liquid-like_2017}%
	\BibitemOpen
	\bibfield  {author} {\bibinfo {author} {\bibfnamefont {J.}~\bibnamefont
			{Wind}}, \bibinfo {author} {\bibfnamefont {R.~A.}\ \bibnamefont {Mole}},
		\bibinfo {author} {\bibfnamefont {D.}~\bibnamefont {Yu}},\ and\ \bibinfo
		{author} {\bibfnamefont {C.~D.}\ \bibnamefont {Ling}},\ }\bibfield  {title}
	{\bibinfo {title} {Liquid-like ionic diffusion in solid bismuth oxide
			revealed by coherent quasielastic neutron scattering},\ }\href
	{https://doi.org/10.1021/acs.chemmater.7b02374} {\bibfield  {journal}
		{\bibinfo  {journal} {Chem. Mater.}\ }\textbf {\bibinfo {volume} {29}},\
		\bibinfo {pages} {7408} (\bibinfo {year} {2017})}\BibitemShut {NoStop}%
	\bibitem [{\citenamefont {Harwig}\ and\ \citenamefont
		{Gerards}(1979)}]{Harwig1979121}%
	\BibitemOpen
	\bibfield  {author} {\bibinfo {author} {\bibfnamefont {H.}~\bibnamefont
			{Harwig}}\ and\ \bibinfo {author} {\bibfnamefont {A.}~\bibnamefont
			{Gerards}},\ }\bibfield  {title} {\bibinfo {title} {The polymorphism of
			bismuth sesquioxide},\ }\href
	{https://doi.org/https://doi.org/10.1016/0040-6031(79)87011-2} {\bibfield
		{journal} {\bibinfo  {journal} {Thermochimica Acta}\ }\textbf {\bibinfo
			{volume} {28}},\ \bibinfo {pages} {121} (\bibinfo {year} {1979})}\BibitemShut
	{NoStop}%
	\bibitem [{\citenamefont {Yashima}\ \emph {et~al.}(2005)\citenamefont
		{Yashima}, \citenamefont {Ishimura},\ and\ \citenamefont
		{Ohoyama}}]{YashimaNPD}%
	\BibitemOpen
	\bibfield  {author} {\bibinfo {author} {\bibfnamefont {M.}~\bibnamefont
			{Yashima}}, \bibinfo {author} {\bibfnamefont {D.}~\bibnamefont {Ishimura}},\
		and\ \bibinfo {author} {\bibfnamefont {K.}~\bibnamefont {Ohoyama}},\
	}\bibfield  {title} {\bibinfo {title} {Temperature dependence of lattice
			parameters and anisotropic thermal expansion of bismuth oxide},\ }\href
	{https://doi.org/https://doi.org/10.1111/j.1551-2916.2005.00432.x} {\bibfield
		{journal} {\bibinfo  {journal} {J. Am. Ceram. Soc.}\ }\textbf {\bibinfo
			{volume} {88}},\ \bibinfo {pages} {2332} (\bibinfo {year}
		{2005})}\BibitemShut {NoStop}%
	\bibitem [{\citenamefont {Harwig}\ and\ \citenamefont
		{Weenk}(1978)}]{19784440119}%
	\BibitemOpen
	\bibfield  {author} {\bibinfo {author} {\bibfnamefont {H.~A.}\ \bibnamefont
			{Harwig}}\ and\ \bibinfo {author} {\bibfnamefont {J.~W.}\ \bibnamefont
			{Weenk}},\ }\bibfield  {title} {\bibinfo {title} {Phase relations in
			bismuthsesquioxide},\ }\href
	{https://doi.org/https://doi.org/10.1002/zaac.19784440119} {\bibfield
		{journal} {\bibinfo  {journal} {Z. anorg. allg. Chem.}\ }\textbf {\bibinfo
			{volume} {444}},\ \bibinfo {pages} {167} (\bibinfo {year}
		{1978})}\BibitemShut {NoStop}%
	\bibitem [{\citenamefont {Yashima}\ \emph {et~al.}(1996)\citenamefont
		{Yashima}, \citenamefont {Ohtake}, \citenamefont {Kakihana}, \citenamefont
		{Arashi},\ and\ \citenamefont {Yoshimura}}]{Yashima1996}%
	\BibitemOpen
	\bibfield  {author} {\bibinfo {author} {\bibfnamefont {M.}~\bibnamefont
			{Yashima}}, \bibinfo {author} {\bibfnamefont {K.}~\bibnamefont {Ohtake}},
		\bibinfo {author} {\bibfnamefont {M.}~\bibnamefont {Kakihana}}, \bibinfo
		{author} {\bibfnamefont {H.}~\bibnamefont {Arashi}},\ and\ \bibinfo {author}
		{\bibfnamefont {M.}~\bibnamefont {Yoshimura}},\ }\href
	{https://doi.org/10.1016/0022-3697(95)00085-2} {\bibfield  {journal}
		{\bibinfo  {journal} {J. Phys. Chem. Solids}\ }\textbf {\bibinfo {volume}
			{57}},\ \bibinfo {pages} {17} (\bibinfo {year} {1996})}\BibitemShut {NoStop}%
	\bibitem [{\citenamefont {Mohn}\ \emph
		{et~al.}(2009{\natexlab{b}})\citenamefont {Mohn}, \citenamefont {St{\o}len},
		\citenamefont {Norberg},\ and\ \citenamefont {Hull}}]{Mohn2009}%
	\BibitemOpen
	\bibfield  {author} {\bibinfo {author} {\bibfnamefont {C.~E.}\ \bibnamefont
			{Mohn}}, \bibinfo {author} {\bibfnamefont {S.}~\bibnamefont {St{\o}len}},
		\bibinfo {author} {\bibfnamefont {S.~T.}\ \bibnamefont {Norberg}},\ and\
		\bibinfo {author} {\bibfnamefont {S.}~\bibnamefont {Hull}},\ }\bibfield
	{title} {\bibinfo {title} {{Ab initio molecular dynamics simulations of
				oxide-ion disorder in the $\delta$ -Bi2 O3}},\ }\href
	{https://doi.org/10.1103/PhysRevB.80.024205} {\bibfield  {journal} {\bibinfo
			{journal} {Phy. Rev. B}\ }\textbf {\bibinfo {volume} {80}},\ \bibinfo {pages}
		{1} (\bibinfo {year} {2009}{\natexlab{b}})}\BibitemShut {NoStop}%
	\bibitem [{\citenamefont {Aidhy}\ \emph {et~al.}(2008)\citenamefont {Aidhy},
		\citenamefont {Nino}, \citenamefont {Sinnott}, \citenamefont {Wachsman},\
		and\ \citenamefont {Phillpotw}}]{Aidhy2008}%
	\BibitemOpen
	\bibfield  {author} {\bibinfo {author} {\bibfnamefont {D.~S.}\ \bibnamefont
			{Aidhy}}, \bibinfo {author} {\bibfnamefont {J.~C.}\ \bibnamefont {Nino}},
		\bibinfo {author} {\bibfnamefont {S.~B.}\ \bibnamefont {Sinnott}}, \bibinfo
		{author} {\bibfnamefont {E.~D.}\ \bibnamefont {Wachsman}},\ and\ \bibinfo
		{author} {\bibfnamefont {S.~R.}\ \bibnamefont {Phillpotw}},\ }\bibfield
	{title} {\bibinfo {title} {{Vacancy-ordered structure of cubic bismuth oxide
				from simulation and crystallographic analysis}},\ }\href
	{https://doi.org/10.1111/j.1551-2916.2008.02463.x} {\bibfield  {journal}
		{\bibinfo  {journal} {J. Am. Ceram. Soc.}\ }\textbf {\bibinfo {volume}
			{91}},\ \bibinfo {pages} {2349} (\bibinfo {year} {2008})}\BibitemShut
	{NoStop}%
	\bibitem [{\citenamefont {Goel}\ \emph {et~al.}(2008)\citenamefont {Goel},
		\citenamefont {Gupta}, \citenamefont {Mittal}, \citenamefont {Skinner},
		\citenamefont {Mukhopadhyay}, \citenamefont {Rols},\ and\ \citenamefont
		{Chaplot}}]{Goel2020}%
	\BibitemOpen
	\bibfield  {author} {\bibinfo {author} {\bibfnamefont {P.}~\bibnamefont
			{Goel}}, \bibinfo {author} {\bibfnamefont {M.~K.}\ \bibnamefont {Gupta}},
		\bibinfo {author} {\bibfnamefont {R.}~\bibnamefont {Mittal}}, \bibinfo
		{author} {\bibfnamefont {S.~J.}\ \bibnamefont {Skinner}}, \bibinfo {author}
		{\bibfnamefont {S.}~\bibnamefont {Mukhopadhyay}}, \bibinfo {author}
		{\bibfnamefont {S.}~\bibnamefont {Rols}},\ and\ \bibinfo {author}
		{\bibfnamefont {S.~L.}\ \bibnamefont {Chaplot}},\ }\href@noop {} {\bibfield
		{journal} {\bibinfo  {journal} {J. Phys.: Condens. Matter}\ }\textbf
		{\bibinfo {volume} {334002}},\ \bibinfo {pages} {1} (\bibinfo {year}
		{2008})}\BibitemShut {NoStop}%
	\bibitem [{\citenamefont {Walsh}\ \emph {et~al.}(2006)\citenamefont {Walsh},
		\citenamefont {Watson}, \citenamefont {Payne}, \citenamefont {Edgell},
		\citenamefont {Guo}, \citenamefont {Glans}, \citenamefont {Learmonth},\ and\
		\citenamefont {Smith}}]{Walsh2006}%
	\BibitemOpen
	\bibfield  {author} {\bibinfo {author} {\bibfnamefont {A.}~\bibnamefont
			{Walsh}}, \bibinfo {author} {\bibfnamefont {G.~W.}\ \bibnamefont {Watson}},
		\bibinfo {author} {\bibfnamefont {D.~J.}\ \bibnamefont {Payne}}, \bibinfo
		{author} {\bibfnamefont {R.~G.}\ \bibnamefont {Edgell}}, \bibinfo {author}
		{\bibfnamefont {J.}~\bibnamefont {Guo}}, \bibinfo {author} {\bibfnamefont
			{P.~A.}\ \bibnamefont {Glans}}, \bibinfo {author} {\bibfnamefont
			{T.}~\bibnamefont {Learmonth}},\ and\ \bibinfo {author} {\bibfnamefont
			{K.~E.}\ \bibnamefont {Smith}},\ }\bibfield  {title} {\bibinfo {title}
		{{Electronic structure of the $\alpha$ and $\delta$ phases of Bi2 O3: A
				combined ab initio and x-ray spectroscopy study}},\ }\href
	{https://doi.org/10.1103/PhysRevB.73.235104} {\bibfield  {journal} {\bibinfo
			{journal} {Phys. Rev. B}\ }\textbf {\bibinfo {volume} {73}},\ \bibinfo
		{pages} {1} (\bibinfo {year} {2006})}\BibitemShut {NoStop}%
	\bibitem [{\citenamefont {Fan}\ \emph {et~al.}(2006)\citenamefont {Fan},
		\citenamefont {Pan}, \citenamefont {Teng}, \citenamefont {Ye},\ and\
		\citenamefont {Li}}]{Fan_2006}%
	\BibitemOpen
	\bibfield  {author} {\bibinfo {author} {\bibfnamefont {H.~T.}\ \bibnamefont
			{Fan}}, \bibinfo {author} {\bibfnamefont {S.~S.}\ \bibnamefont {Pan}},
		\bibinfo {author} {\bibfnamefont {X.~M.}\ \bibnamefont {Teng}}, \bibinfo
		{author} {\bibfnamefont {C.}~\bibnamefont {Ye}},\ and\ \bibinfo {author}
		{\bibfnamefont {G.~H.}\ \bibnamefont {Li}},\ }\bibfield  {title} {\bibinfo
		{title} {},\ }\href
	{https://doi.org/10.1088/0022-3727/39/9/032} {\bibfield  {journal} {\bibinfo
			{journal} {J. Phys. D: Appl. Phys}\ }\textbf {\bibinfo {volume} {39}},\
		\bibinfo {pages} {1939} (\bibinfo {year} {2006})}\BibitemShut {NoStop}%
	\bibitem [{\citenamefont {Matsumoto}\ \emph {et~al.}(2010)\citenamefont
		{Matsumoto}, \citenamefont {Koyama},\ and\ \citenamefont
		{Tanaka}}]{Matsumoto2010}%
	\BibitemOpen
	\bibfield  {author} {\bibinfo {author} {\bibfnamefont {A.}~\bibnamefont
			{Matsumoto}}, \bibinfo {author} {\bibfnamefont {Y.}~\bibnamefont {Koyama}},\
		and\ \bibinfo {author} {\bibfnamefont {I.}~\bibnamefont {Tanaka}},\
	}\bibfield  {title} {\bibinfo {title} {{Structures and energetics of Bi2 O3
				polymorphs in a defective fluorite family derived by systematic
				first-principles lattice dynamics calculations}},\ }\href
	{https://doi.org/10.1103/PhysRevB.81.094117} {\bibfield  {journal} {\bibinfo
			{journal} {Phys. Rev. B}\ }\textbf {\bibinfo {volume} {81}},\ \bibinfo
		{pages} {1} (\bibinfo {year} {2010})}\BibitemShut {NoStop}%
	\bibitem [{\citenamefont {Schr{\"{o}}der}\ \emph {et~al.}(2010)\citenamefont
		{Schr{\"{o}}der}, \citenamefont {Bagdassarov}, \citenamefont {Ritter},\ and\
		\citenamefont {Bayarjargal}}]{Schroder2010}%
	\BibitemOpen
	\bibfield  {author} {\bibinfo {author} {\bibfnamefont {F.}~\bibnamefont
			{Schr{\"{o}}der}}, \bibinfo {author} {\bibfnamefont {N.}~\bibnamefont
			{Bagdassarov}}, \bibinfo {author} {\bibfnamefont {F.}~\bibnamefont
			{Ritter}},\ and\ \bibinfo {author} {\bibfnamefont {L.}~\bibnamefont
			{Bayarjargal}},\ }\bibfield  {title} {\bibinfo {title} {{Temperature
				dependence of Bi2O3 structural parameters close to the $\alpha$–$\delta$
				phase transition}},\ }\href {https://doi.org/10.1080/01411591003795290}
	{\bibfield  {journal} {\bibinfo  {journal} {Phase Transitions}\ }\textbf
		{\bibinfo {volume} {83}},\ \bibinfo {pages} {311} (\bibinfo {year}
		{2010})}\BibitemShut {NoStop}%
	\bibitem [{\citenamefont {Zhu}\ \emph {et~al.}(2016)\citenamefont {Zhu},
		\citenamefont {An}, \citenamefont {Yu}, \citenamefont {Marcelli},
		\citenamefont {Liu}, \citenamefont {Hu},\ and\ \citenamefont
		{Xu}}]{Zhu_2016}%
	\BibitemOpen
	\bibfield  {author} {\bibinfo {author} {\bibfnamefont {Y.}~\bibnamefont
			{Zhu}}, \bibinfo {author} {\bibfnamefont {P.}~\bibnamefont {An}}, \bibinfo
		{author} {\bibfnamefont {M.}~\bibnamefont {Yu}}, \bibinfo {author}
		{\bibfnamefont {A.}~\bibnamefont {Marcelli}}, \bibinfo {author}
		{\bibfnamefont {Y.}~\bibnamefont {Liu}}, \bibinfo {author} {\bibfnamefont
			{T.}~\bibnamefont {Hu}},\ and\ \bibinfo {author} {\bibfnamefont
			{W.}~\bibnamefont {Xu}},\ }\bibfield  {title} {\bibinfo {title} {Structural
			phase transitions in ionic conductor bi2o3by temperature dependent {XPD} and
			{XAS}},\ }\href {https://doi.org/10.1088/1742-6596/712/1/012132} {\bibfield
		{journal} {\bibinfo  {journal} {Journal of Physics: Conference Series}\
		}\textbf {\bibinfo {volume} {712}},\ \bibinfo {pages} {012132} (\bibinfo
		{year} {2016})}\BibitemShut {NoStop}%
	\bibitem [{\citenamefont {Klinkova}\ \emph {et~al.}(2007)\citenamefont
		{Klinkova}, \citenamefont {Nikolaichik}, \citenamefont {Barkovskii},\ and\
		\citenamefont {Fedotov}}]{Klinkova2007}%
	\BibitemOpen
	\bibfield  {author} {\bibinfo {author} {\bibfnamefont {L.~A.}\ \bibnamefont
			{Klinkova}}, \bibinfo {author} {\bibfnamefont {V.~I.}\ \bibnamefont
			{Nikolaichik}}, \bibinfo {author} {\bibfnamefont {N.~V.}\ \bibnamefont
			{Barkovskii}},\ and\ \bibinfo {author} {\bibfnamefont {V.~K.}\ \bibnamefont
			{Fedotov}},\ }\bibfield  {title} {\bibinfo {title} {{Thermal stability of
				Bi2O3}},\ }\href {https://doi.org/10.1134/S0036023607120030} {\bibfield
		{journal} {\bibinfo  {journal} {Russ. J. Inorg. Chem.}\ }\textbf {\bibinfo
			{volume} {52}},\ \bibinfo {pages} {1822} (\bibinfo {year}
		{2007})}\BibitemShut {NoStop}%
	\bibitem [{SI()}]{SI}%
	\BibitemOpen
	\href@noop {} {\bibinfo  {journal} {See Supporting information at
			http://link.aps.org for supporting experimental and data fitting details}\
	}\BibitemShut {NoStop}%
	\bibitem [{\citenamefont {Pereira}\ \emph {et~al.}(2014)\citenamefont
		{Pereira}, \citenamefont {Gomis}, \citenamefont {Sans}, \citenamefont
		{Pellicer-Porres}, \citenamefont {Manj{\'{o}}n}, \citenamefont {Beltran},
		\citenamefont {Rodr{\'{\i}}guez-Hern{\'{a}}ndez},\ and\ \citenamefont
		{Mu{\~{n}}oz}}]{Pereira_2014}%
	\BibitemOpen
	\bibfield  {journal} {  }\bibfield  {author} {\bibinfo {author} {\bibfnamefont
			{A.~L.~J.}\ \bibnamefont {Pereira}}, \bibinfo {author} {\bibfnamefont
			{O.}~\bibnamefont {Gomis}}, \bibinfo {author} {\bibfnamefont {J.~A.}\
			\bibnamefont {Sans}}, \bibinfo {author} {\bibfnamefont {J.}~\bibnamefont
			{Pellicer-Porres}}, \bibinfo {author} {\bibfnamefont {F.~J.}\ \bibnamefont
			{Manj{\'{o}}n}}, \bibinfo {author} {\bibfnamefont {A.}~\bibnamefont
			{Beltran}}, \bibinfo {author} {\bibfnamefont {P.}~\bibnamefont
			{Rodr{\'{\i}}guez-Hern{\'{a}}ndez}},\ and\ \bibinfo {author} {\bibfnamefont
			{A.}~\bibnamefont {Mu{\~{n}}oz}},\ }\bibfield  {title} {\bibinfo {title}
		{},\ }\href
	{https://doi.org/10.1088/0953-8984/26/22/225401} {\bibfield  {journal}
		{\bibinfo  {journal} {J. Phys.: Condens. Matter}\ }\textbf {\bibinfo {volume}
			{26}},\ \bibinfo {pages} {225401} (\bibinfo {year} {2014})}\BibitemShut
	{NoStop}%
	\bibitem [{\citenamefont {Betsch}\ and\ \citenamefont
		{White}(1978)}]{BETSCH1978505}%
	\BibitemOpen
	\bibfield  {author} {\bibinfo {author} {\bibfnamefont {R.~J.}\ \bibnamefont
			{Betsch}}\ and\ \bibinfo {author} {\bibfnamefont {W.~B.}\ \bibnamefont
			{White}},\ }\bibfield  {title} {\bibinfo {title} {Vibrational spectra of
			bismuth oxide and the sillenite-structure bismuth oxide derivatives},\ }\href
	{https://doi.org/https://doi.org/10.1016/0584-8539(78)80047-6} {\bibfield
		{journal} {\bibinfo  {journal} {Spectrochim. Acta A Mol. Biomol. Spectrosc.}\
		}\textbf {\bibinfo {volume} {34}},\ \bibinfo {pages} {505} (\bibinfo {year}
		{1978})}\BibitemShut {NoStop}%
	\bibitem [{\citenamefont {Denisov}\ \emph {et~al.}(1997)\citenamefont
		{Denisov}, \citenamefont {Ivlev}, \citenamefont {Lipin}, \citenamefont
		{Mavrin},\ and\ \citenamefont {Orlov}}]{Denisov1997}%
	\BibitemOpen
	\bibfield  {author} {\bibinfo {author} {\bibfnamefont {V.~N.}\ \bibnamefont
			{Denisov}}, \bibinfo {author} {\bibfnamefont {A.~N.}\ \bibnamefont {Ivlev}},
		\bibinfo {author} {\bibfnamefont {A.~S.}\ \bibnamefont {Lipin}}, \bibinfo
		{author} {\bibfnamefont {B.~N.}\ \bibnamefont {Mavrin}},\ and\ \bibinfo
		{author} {\bibfnamefont {V.~G.}\ \bibnamefont {Orlov}},\ }\bibfield  {title}
	{\bibinfo {title} {{Raman spectra and lattice dynamics of single-crystal}},\
	}\href {https://doi.org/10.1088/0953-8984/9/23/020} {\bibfield  {journal}
		{\bibinfo  {journal} {J. Phys.: Condens. Matter}\ }\textbf {\bibinfo {volume}
			{9}},\ \bibinfo {pages} {4967} (\bibinfo {year} {1997})}\BibitemShut
	{NoStop}%
	\bibitem [{\citenamefont {Cardona}\ \emph {et~al.}(1982)\citenamefont
		{Cardona}, \citenamefont {Chang}, \citenamefont {G{\"{u}}ntherodt},
		\citenamefont {Long},\ and\ \citenamefont {Vogt}}]{Cardona1982}%
	\BibitemOpen
	\bibfield  {author} {\bibinfo {author} {\bibfnamefont {M.}~\bibnamefont
			{Cardona}}, \bibinfo {author} {\bibfnamefont {R.}~\bibnamefont {Chang}},
		\bibinfo {author} {\bibfnamefont {G.}~\bibnamefont {G{\"{u}}ntherodt}},
		\bibinfo {author} {\bibfnamefont {M.}~\bibnamefont {Long}},\ and\ \bibinfo
		{author} {\bibfnamefont {H.}~\bibnamefont {Vogt}},\ }\href
	{https://doi.org/10.1007/3-540-11380-0} {\emph {\bibinfo {title} {{Light
					Scattering in Solids II}}}},\ edited by\ \bibinfo {editor} {\bibfnamefont
		{M.}~\bibnamefont {Cardona}}\ and\ \bibinfo {editor} {\bibfnamefont
		{G.}~\bibnamefont {G{\"{u}}ntherodt}},\ \bibinfo {series} {Topics in Applied
		Physics}, Vol.~\bibinfo {volume} {50}\ (\bibinfo  {publisher} {Springer
		Berlin Heidelberg},\ \bibinfo {address} {Berlin, Heidelberg},\ \bibinfo
	{year} {1982})\ p.\ \bibinfo {pages} {254}\BibitemShut {NoStop}%
	\bibitem [{\citenamefont {Benshalom}\ \emph {et~al.}(2022)\citenamefont
		{Benshalom}, \citenamefont {Reuveni}, \citenamefont {Korobko}, \citenamefont
		{Yaffe},\ and\ \citenamefont {Hellman}}]{Nimrod}%
	\BibitemOpen
	\bibfield  {author} {\bibinfo {author} {\bibfnamefont {N.}~\bibnamefont
			{Benshalom}}, \bibinfo {author} {\bibfnamefont {G.}~\bibnamefont {Reuveni}},
		\bibinfo {author} {\bibfnamefont {R.}~\bibnamefont {Korobko}}, \bibinfo
		{author} {\bibfnamefont {O.}~\bibnamefont {Yaffe}},\ and\ \bibinfo {author}
		{\bibfnamefont {O.}~\bibnamefont {Hellman}},\ }\bibfield  {title} {\bibinfo
		{title} {Dielectric response of rock-salt crystals at finite temperatures
			from first principles},\ }\href
	{https://doi.org/10.1103/PhysRevMaterials.6.033607} {\bibfield  {journal}
		{\bibinfo  {journal} {Phys. Rev. Materials}\ }\textbf {\bibinfo {volume}
			{6}},\ \bibinfo {pages} {033607} (\bibinfo {year} {2022})}\BibitemShut
	{NoStop}%
	\bibitem [{\citenamefont {Schmitt}\ \emph {et~al.}(2020)\citenamefont
		{Schmitt}, \citenamefont {Nenning}, \citenamefont {Kraynis}, \citenamefont
		{Korobko}, \citenamefont {I.~Frenkel}, \citenamefont {Lubomirsky},
		\citenamefont {M.~Haile},\ and\ \citenamefont
		{M.~Rupp}}]{schmitt_review_2020}%
	\BibitemOpen
	\bibfield  {author} {\bibinfo {author} {\bibfnamefont {R.}~\bibnamefont
			{Schmitt}}, \bibinfo {author} {\bibfnamefont {A.}~\bibnamefont {Nenning}},
		\bibinfo {author} {\bibfnamefont {O.}~\bibnamefont {Kraynis}}, \bibinfo
		{author} {\bibfnamefont {R.}~\bibnamefont {Korobko}}, \bibinfo {author}
		{\bibfnamefont {A.}~\bibnamefont {I.~Frenkel}}, \bibinfo {author}
		{\bibfnamefont {I.}~\bibnamefont {Lubomirsky}}, \bibinfo {author}
		{\bibfnamefont {S.}~\bibnamefont {M.~Haile}},\ and\ \bibinfo {author}
		{\bibfnamefont {J.~L.}\ \bibnamefont {M.~Rupp}},\ }\bibfield  {title}
	{\bibinfo {title} {A review of defect structure and chemistry in ceria and
			its solid solutions},\ }\bibfield  {journal} {\bibinfo  {journal} {Chem. Soc.
			Rev.}\ }\href {https://doi.org/10.1039/C9CS00588A} {10.1039/C9CS00588A}
	(\bibinfo {year} {2020})\BibitemShut {NoStop}%
	\bibitem [{\citenamefont {{Fernandez Lopez}}\ \emph {et~al.}(2001)\citenamefont
		{{Fernandez Lopez}}, \citenamefont {{Sanchez Escribano}}, \citenamefont
		{Panizza}, \citenamefont {Carnasciali},\ and\ \citenamefont
		{Busca}}]{FernandezLopez2001}%
	\BibitemOpen
	\bibfield  {author} {\bibinfo {author} {\bibfnamefont {E.}~\bibnamefont
			{{Fernandez Lopez}}}, \bibinfo {author} {\bibfnamefont {V.}~\bibnamefont
			{{Sanchez Escribano}}}, \bibinfo {author} {\bibfnamefont {M.}~\bibnamefont
			{Panizza}}, \bibinfo {author} {\bibfnamefont {M.~M.}\ \bibnamefont
			{Carnasciali}},\ and\ \bibinfo {author} {\bibfnamefont {G.}~\bibnamefont
			{Busca}},\ }\bibfield  {title} {\bibinfo {title} {{Vibrational and electronic
				spectroscopic properties of zirconia powders}},\ }\href
	{https://doi.org/10.1039/b100909p} {\bibfield  {journal} {\bibinfo  {journal}
			{J. Mat. Chem.}\ }\textbf {\bibinfo {volume} {11}},\ \bibinfo {pages} {1891}
		(\bibinfo {year} {2001})}\BibitemShut {NoStop}%
	\bibitem [{\citenamefont {Gattow}\ and\ \citenamefont
		{Schr{\"{o}}der}(1962)}]{Gattow}%
	\BibitemOpen
	\bibfield  {author} {\bibinfo {author} {\bibfnamefont {G.}~\bibnamefont
			{Gattow}}\ and\ \bibinfo {author} {\bibfnamefont {H.}~\bibnamefont
			{Schr{\"{o}}der}},\ }\bibfield  {title} {\bibinfo {title} {{{\"{U}}ber
				Wismutoxide. III. Die Kristallstruktur der Hochtemperaturmodifikation von
				Wismut(III)-oxid ($\delta$-Bi2O3)}},\ }\href
	{https://doi.org/https://doi.org/10.1002/zaac.19623180307} {\bibfield
		{journal} {\bibinfo  {journal} {Z. Anorg. Allg. Chem.}\ }\textbf {\bibinfo
			{volume} {318}},\ \bibinfo {pages} {176} (\bibinfo {year}
		{1962})}\BibitemShut {NoStop}%
	\bibitem [{\citenamefont {Harwig}(1978)}]{Harwig}%
	\BibitemOpen
	\bibfield  {author} {\bibinfo {author} {\bibfnamefont {H.~A.}\ \bibnamefont
			{Harwig}},\ }\href {https://doi.org/https://doi.org/10.1002/zaac.19784440118}
	{\bibfield  {journal} {\bibinfo  {journal} {Z. Anorg. Allg. Chem.}\ }\textbf
		{\bibinfo {volume} {444}},\ \bibinfo {pages} {151} (\bibinfo {year}
		{1978})}\BibitemShut {NoStop}%
	\bibitem [{\citenamefont {Armstrong}(1989)}]{Armstrong1989}%
	\BibitemOpen
	\bibfield  {author} {\bibinfo {author} {\bibfnamefont {R.~L.}\ \bibnamefont
			{Armstrong}},\ }\bibfield  {title} {\bibinfo {title} {Displacive
			order—disorder crossover in perovskite and antifluorite crystals undergoing
			rotational phase transitions},\ }\href
	{https://doi.org/https://doi.org/10.1016/0079-6565(89)80002-0} {\bibfield
		{journal} {\bibinfo  {journal} {Prog. Nucl. Magn. Reson. Spectrosc.}\
		}\textbf {\bibinfo {volume} {21}},\ \bibinfo {pages} {151} (\bibinfo {year}
		{1989})}\BibitemShut {NoStop}%
	\bibitem [{\citenamefont {Roleder}\ \emph {et~al.}(2000)\citenamefont
		{Roleder}, \citenamefont {Jankowska-Sumara}, \citenamefont {Kugel},
		\citenamefont {Maglione}, \citenamefont {Fontana},\ and\ \citenamefont
		{Dec}}]{Roleder2000}%
	\BibitemOpen
	\bibfield  {author} {\bibinfo {author} {\bibfnamefont {K.}~\bibnamefont
			{Roleder}}, \bibinfo {author} {\bibfnamefont {I.}~\bibnamefont
			{Jankowska-Sumara}}, \bibinfo {author} {\bibfnamefont {G.~E.}\ \bibnamefont
			{Kugel}}, \bibinfo {author} {\bibfnamefont {M.}~\bibnamefont {Maglione}},
		\bibinfo {author} {\bibfnamefont {M.~D.}\ \bibnamefont {Fontana}},\ and\
		\bibinfo {author} {\bibfnamefont {J.}~\bibnamefont {Dec}},\ }\bibfield
	{title} {\bibinfo {title} {{Antiferroelectric and ferroelectric phase
				transitions of the displacive and order-disorder type in PbZrO3 and PbZr1-xTi
				x O3 single crystals}},\ }\href {https://doi.org/10.1080/01411590008209310}
	{\bibfield  {journal} {\bibinfo  {journal} {Phase Transitions}\ }\textbf
		{\bibinfo {volume} {71}},\ \bibinfo {pages} {287} (\bibinfo {year}
		{2000})}\BibitemShut {NoStop}%
	\bibitem [{\citenamefont {Xu}\ \emph {et~al.}(2019)\citenamefont {Xu},
		\citenamefont {Hellman},\ and\ \citenamefont {Bellaiche}}]{Xu2019}%
	\BibitemOpen
	\bibfield  {author} {\bibinfo {author} {\bibfnamefont {B.}~\bibnamefont
			{Xu}}, \bibinfo {author} {\bibfnamefont {O.}~\bibnamefont {Hellman}},\ and\
		\bibinfo {author} {\bibfnamefont {L.}~\bibnamefont {Bellaiche}},\ }\bibfield
	{title} {\bibinfo {title} {Order-disorder transition in the prototypical
			antiferroelectric ${\mathrm{pbzro}}_{3}$},\ }\href
	{https://doi.org/10.1103/PhysRevB.100.020102} {\bibfield  {journal} {\bibinfo
			{journal} {Phys. Rev. B}\ }\textbf {\bibinfo {volume} {100}},\ \bibinfo
		{pages} {020102} (\bibinfo {year} {2019})}\BibitemShut {NoStop}%
	\bibitem [{\citenamefont {Scott}(1974)}]{Scott1974}%
	\BibitemOpen
	\bibfield  {author} {\bibinfo {author} {\bibfnamefont {J.~F.}\ \bibnamefont
			{Scott}},\ }\bibfield  {title} {\bibinfo {title} {Soft-mode spectroscopy:
			Experimental studies of structural phase transitions},\ }\href
	{https://doi.org/10.1103/RevModPhys.46.83} {\bibfield  {journal} {\bibinfo
			{journal} {Rev. Mod. Phys.}\ }\textbf {\bibinfo {volume} {46}},\ \bibinfo
		{pages} {83} (\bibinfo {year} {1974})}\BibitemShut {NoStop}%
	\bibitem [{\citenamefont {Yaffe}\ \emph {et~al.}(2017)\citenamefont {Yaffe},
		\citenamefont {Guo}, \citenamefont {Tan}, \citenamefont {Egger},
		\citenamefont {Hull}, \citenamefont {Stoumpos}, \citenamefont {Zheng},
		\citenamefont {Heinz}, \citenamefont {Kronik}, \citenamefont {Kanatzidis},
		\citenamefont {Owen}, \citenamefont {Rappe}, \citenamefont {Pimenta},\ and\
		\citenamefont {Brus}}]{PhysRevLett.118.136001}%
	\BibitemOpen
	\bibfield  {author} {\bibinfo {author} {\bibfnamefont {O.}~\bibnamefont
			{Yaffe}}, \bibinfo {author} {\bibfnamefont {Y.}~\bibnamefont {Guo}}, \bibinfo
		{author} {\bibfnamefont {L.~Z.}\ \bibnamefont {Tan}}, \bibinfo {author}
		{\bibfnamefont {D.~A.}\ \bibnamefont {Egger}}, \bibinfo {author}
		{\bibfnamefont {T.}~\bibnamefont {Hull}}, \bibinfo {author} {\bibfnamefont
			{C.~C.}\ \bibnamefont {Stoumpos}}, \bibinfo {author} {\bibfnamefont
			{F.}~\bibnamefont {Zheng}}, \bibinfo {author} {\bibfnamefont {T.~F.}\
			\bibnamefont {Heinz}}, \bibinfo {author} {\bibfnamefont {L.}~\bibnamefont
			{Kronik}}, \bibinfo {author} {\bibfnamefont {M.~G.}\ \bibnamefont
			{Kanatzidis}}, \bibinfo {author} {\bibfnamefont {J.~S.}\ \bibnamefont
			{Owen}}, \bibinfo {author} {\bibfnamefont {A.~M.}\ \bibnamefont {Rappe}},
		\bibinfo {author} {\bibfnamefont {M.~A.}\ \bibnamefont {Pimenta}},\ and\
		\bibinfo {author} {\bibfnamefont {L.~E.}\ \bibnamefont {Brus}},\ }\bibfield
	{title} {\bibinfo {title} {Local polar fluctuations in lead halide perovskite
			crystals},\ }\href {https://doi.org/10.1103/PhysRevLett.118.136001}
	{\bibfield  {journal} {\bibinfo  {journal} {Phys. Rev. Lett.}\ }\textbf
		{\bibinfo {volume} {118}},\ \bibinfo {pages} {136001} (\bibinfo {year}
		{2017})}\BibitemShut {NoStop}%
	\bibitem [{\citenamefont {Beecher}\ \emph {et~al.}(2016)\citenamefont
		{Beecher}, \citenamefont {Semonin}, \citenamefont {Skelton}, \citenamefont
		{Frost}, \citenamefont {Terban}, \citenamefont {Zhai}, \citenamefont
		{Alatas}, \citenamefont {Owen}, \citenamefont {Walsh},\ and\ \citenamefont
		{Billinge}}]{doi:10.1021/acsenergylett.6b00381}%
	\BibitemOpen
	\bibfield  {author} {\bibinfo {author} {\bibfnamefont {A.~N.}\ \bibnamefont
			{Beecher}}, \bibinfo {author} {\bibfnamefont {O.~E.}\ \bibnamefont
			{Semonin}}, \bibinfo {author} {\bibfnamefont {J.~M.}\ \bibnamefont
			{Skelton}}, \bibinfo {author} {\bibfnamefont {J.~M.}\ \bibnamefont {Frost}},
		\bibinfo {author} {\bibfnamefont {M.~W.}\ \bibnamefont {Terban}}, \bibinfo
		{author} {\bibfnamefont {H.}~\bibnamefont {Zhai}}, \bibinfo {author}
		{\bibfnamefont {A.}~\bibnamefont {Alatas}}, \bibinfo {author} {\bibfnamefont
			{J.~S.}\ \bibnamefont {Owen}}, \bibinfo {author} {\bibfnamefont
			{A.}~\bibnamefont {Walsh}},\ and\ \bibinfo {author} {\bibfnamefont
			{S.~J.~L.}\ \bibnamefont {Billinge}},\ }\bibfield  {title} {\bibinfo {title}
		{{Direct Observation of Dynamic Symmetry Breaking above Room Temperature in
				Methylammonium Lead Iodide Perovskite}},\ }\href
	{https://doi.org/10.1021/acsenergylett.6b00381} {\bibfield  {journal}
		{\bibinfo  {journal} {ACS Energy Lett.}\ }\textbf {\bibinfo {volume} {1}},\
		\bibinfo {pages} {880} (\bibinfo {year} {2016})}\BibitemShut {NoStop}%
	\bibitem [{\citenamefont {Picozzi}\ \emph {et~al.}(2008)\citenamefont
		{Picozzi}, \citenamefont {Yamauchi}, \citenamefont {Sergienko}, \citenamefont
		{Sen}, \citenamefont {Sanyal},\ and\ \citenamefont {Dagotto}}]{Picozzi_2008}%
	\BibitemOpen
	\bibfield  {author} {\bibinfo {author} {\bibfnamefont {S.}~\bibnamefont
			{Picozzi}}, \bibinfo {author} {\bibfnamefont {K.}~\bibnamefont {Yamauchi}},
		\bibinfo {author} {\bibfnamefont {I.~A.}\ \bibnamefont {Sergienko}}, \bibinfo
		{author} {\bibfnamefont {C.}~\bibnamefont {Sen}}, \bibinfo {author}
		{\bibfnamefont {B.}~\bibnamefont {Sanyal}},\ and\ \bibinfo {author}
		{\bibfnamefont {E.}~\bibnamefont {Dagotto}},\ }\bibfield  {title} {\bibinfo
		{title} {{Microscopic mechanisms for improper ferroelectricity in
				multiferroic perovskites: a theoretical review}},\ }\href
	{https://doi.org/10.1088/0953-8984/20/43/434208} {\bibfield  {journal}
		{\bibinfo  {journal} {J. Phys.: Condens. Matter}\ }\textbf {\bibinfo {volume}
			{20}},\ \bibinfo {pages} {434208} (\bibinfo {year} {2008})}\BibitemShut
	{NoStop}%
	\bibitem [{\citenamefont {Fontana}\ \emph {et~al.}(2020)\citenamefont
		{Fontana}, \citenamefont {Kokanyan},\ and\ \citenamefont
		{Kauffmann}}]{Fontana_2020}%
	\BibitemOpen
	\bibfield  {author} {\bibinfo {author} {\bibfnamefont {M.~D.}\ \bibnamefont
			{Fontana}}, \bibinfo {author} {\bibfnamefont {N.}~\bibnamefont {Kokanyan}},\
		and\ \bibinfo {author} {\bibfnamefont {T.~H.}\ \bibnamefont {Kauffmann}},\
	}\bibfield  {title} {\bibinfo {title} {{}},\ }\href {https://doi.org/10.1088/1361-648x/ab808e}
	{\bibfield  {journal} {\bibinfo  {journal} {J. Phys.: Condens. Matter}\
		}\textbf {\bibinfo {volume} {32}},\ \bibinfo {pages} {285403} (\bibinfo
		{year} {2020})}\BibitemShut {NoStop}%
	\bibitem [{\citenamefont {Sharma}\ \emph {et~al.}(2020)\citenamefont {Sharma},
		\citenamefont {Dai}, \citenamefont {Gao}, \citenamefont {Brenner},
		\citenamefont {Yadgarov}, \citenamefont {Zhang}, \citenamefont {Rakita},
		\citenamefont {Korobko}, \citenamefont {Rappe},\ and\ \citenamefont
		{Yaffe}}]{Sharma}%
	\BibitemOpen
	\bibfield  {author} {\bibinfo {author} {\bibfnamefont {R.}~\bibnamefont
			{Sharma}}, \bibinfo {author} {\bibfnamefont {Z.}~\bibnamefont {Dai}},
		\bibinfo {author} {\bibfnamefont {L.}~\bibnamefont {Gao}}, \bibinfo {author}
		{\bibfnamefont {T.~M.}\ \bibnamefont {Brenner}}, \bibinfo {author}
		{\bibfnamefont {L.}~\bibnamefont {Yadgarov}}, \bibinfo {author}
		{\bibfnamefont {J.}~\bibnamefont {Zhang}}, \bibinfo {author} {\bibfnamefont
			{Y.}~\bibnamefont {Rakita}}, \bibinfo {author} {\bibfnamefont
			{R.}~\bibnamefont {Korobko}}, \bibinfo {author} {\bibfnamefont {A.~M.}\
			\bibnamefont {Rappe}},\ and\ \bibinfo {author} {\bibfnamefont
			{O.}~\bibnamefont {Yaffe}},\ }\bibfield  {title} {\bibinfo {title}
		{Elucidating the atomistic origin of anharmonicity in tetragonal
			${\mathrm{ch}}_{3}{\mathrm{nh}}_{3}{\mathrm{pbi}}_{3}$ with raman
			scattering},\ }\href {https://doi.org/10.1103/PhysRevMaterials.4.092401}
	{\bibfield  {journal} {\bibinfo  {journal} {Phys. Rev. Mater.}\ }\textbf
		{\bibinfo {volume} {4}},\ \bibinfo {pages} {092401} (\bibinfo {year}
		{2020})}\BibitemShut {NoStop}%
	\bibitem [{\citenamefont {Blount}\ \emph {et~al.}(1962)\citenamefont {Blount},
		\citenamefont {Shul-man}, \citenamefont {{Cohen E I Blount}}, \citenamefont
		{Lowdin}, \citenamefont {Pullman}, \citenamefont {Fleury},\ and\
		\citenamefont {Scott}}]{Blount1962}%
	\BibitemOpen
	\bibfield  {author} {\bibinfo {author} {\bibfnamefont {E.~I.}\ \bibnamefont
			{Blount}}, \bibinfo {author} {\bibfnamefont {R.~G.}\ \bibnamefont
			{Shul-man}}, \bibinfo {author} {\bibfnamefont {M.~H.}\ \bibnamefont {{Cohen E
					I Blount}}}, \bibinfo {author} {\bibfnamefont {P.~O.}\ \bibnamefont
			{Lowdin}}, \bibinfo {author} {\bibfnamefont {B.}~\bibnamefont {Pullman}},
		\bibinfo {author} {\bibfnamefont {P.~A.}\ \bibnamefont {Fleury}},\ and\
		\bibinfo {author} {\bibfnamefont {J.~F.}\ \bibnamefont {Scott}},\ }\bibfield
	{title} {\bibinfo {title} {{SOFT PHONON MODES AND THE 110'K PHASE TRANSITION
				IN SrTi03}},\ }\href@noop {} {\bibfield  {journal} {\bibinfo  {journal} {Rev.
				Mod. Phys}\ }\textbf {\bibinfo {volume} {21}},\ \bibinfo {pages} {67}
		(\bibinfo {year} {1962})}\BibitemShut {NoStop}%
	\bibitem [{\citenamefont {Brenner}\ \emph
		{et~al.}(2020{\natexlab{a}})\citenamefont {Brenner}, \citenamefont
		{Gehrmann}, \citenamefont {Korobko}, \citenamefont {Livneh}, \citenamefont
		{Egger},\ and\ \citenamefont {Yaffe}}]{Brenner2020}%
	\BibitemOpen
	\bibfield  {author} {\bibinfo {author} {\bibfnamefont {T.~M.}\ \bibnamefont
			{Brenner}}, \bibinfo {author} {\bibfnamefont {C.}~\bibnamefont {Gehrmann}},
		\bibinfo {author} {\bibfnamefont {R.}~\bibnamefont {Korobko}}, \bibinfo
		{author} {\bibfnamefont {T.}~\bibnamefont {Livneh}}, \bibinfo {author}
		{\bibfnamefont {D.~A.}\ \bibnamefont {Egger}},\ and\ \bibinfo {author}
		{\bibfnamefont {O.}~\bibnamefont {Yaffe}},\ }\bibfield  {title} {\bibinfo
		{title} {{Anharmonic host-lattice dynamics enable fast ion conduction in
				superionic AgI}},\ }\href
	{https://doi.org/10.1103/PHYSREVMATERIALS.4.115402/FIGURES/3/MEDIUM}
	{\bibfield  {journal} {\bibinfo  {journal} {Phys. Rev. Mater.}\ }\textbf
		{\bibinfo {volume} {4}},\ \bibinfo {pages} {115402} (\bibinfo {year}
		{2020}{\natexlab{a}})},\ \Eprint {https://arxiv.org/abs/1911.07492}
	{arXiv:1911.07492} \BibitemShut {NoStop}%
	\bibitem [{\citenamefont {Brenner}\ \emph {et~al.}(2021)\citenamefont
		{Brenner}, \citenamefont {Grumet}, \citenamefont {Till}, \citenamefont
		{Zeier}, \citenamefont {Egger},\ and\ \citenamefont {Yaffe}}]{Brenner2021}%
	\BibitemOpen
	\bibfield  {author} {\bibinfo {author} {\bibfnamefont {T.~M.}\ \bibnamefont
			{Brenner}}, \bibinfo {author} {\bibfnamefont {M.}~\bibnamefont {Grumet}},
		\bibinfo {author} {\bibfnamefont {P.}~\bibnamefont {Till}}, \bibinfo {author}
		{\bibfnamefont {W.~G.}\ \bibnamefont {Zeier}}, \bibinfo {author}
		{\bibfnamefont {D.~A.}\ \bibnamefont {Egger}},\ and\ \bibinfo {author}
		{\bibfnamefont {O.}~\bibnamefont {Yaffe}},\ }\bibfield  {title} {\bibinfo
		{title} {{Anharmonic Lattice Dynamics in Sodium Ion Conductors}},\
	}\href@noop {} {\bibfield  {journal} {\bibinfo  {journal} {Arxiv}\ ,\
			\bibinfo {pages} {1}} (\bibinfo {year} {2021})}\BibitemShut {NoStop}%
	\bibitem [{\citenamefont {Cohen}\ \emph {et~al.}(2022)\citenamefont {Cohen},
		\citenamefont {Brenner}, \citenamefont {Klarbring}, \citenamefont {Sharma},
		\citenamefont {Fabini}, \citenamefont {Korobko}, \citenamefont {Nayak},
		\citenamefont {Hellman},\ and\ \citenamefont {Yaffe}}]{202107932}%
	\BibitemOpen
	\bibfield  {author} {\bibinfo {author} {\bibfnamefont {A.}~\bibnamefont
			{Cohen}}, \bibinfo {author} {\bibfnamefont {T.~M.}\ \bibnamefont {Brenner}},
		\bibinfo {author} {\bibfnamefont {J.}~\bibnamefont {Klarbring}}, \bibinfo
		{author} {\bibfnamefont {R.}~\bibnamefont {Sharma}}, \bibinfo {author}
		{\bibfnamefont {D.~H.}\ \bibnamefont {Fabini}}, \bibinfo {author}
		{\bibfnamefont {R.}~\bibnamefont {Korobko}}, \bibinfo {author} {\bibfnamefont
			{P.~K.}\ \bibnamefont {Nayak}}, \bibinfo {author} {\bibfnamefont
			{O.}~\bibnamefont {Hellman}},\ and\ \bibinfo {author} {\bibfnamefont
			{O.}~\bibnamefont {Yaffe}},\ }\bibfield  {title} {\bibinfo {title}
		{{Diverging Expressions of Anharmonicity in Halide Perovskites}},\ }\href
	{https://doi.org/https://doi.org/10.1002/adma.202107932} {\bibfield
		{journal} {\bibinfo  {journal} {Adv. Mater.}\ }\textbf {\bibinfo {volume}
			{34}},\ \bibinfo {pages} {2107932} (\bibinfo {year} {2022})}\BibitemShut
	{NoStop}%
	\bibitem [{\citenamefont {Menahem}\ \emph {et~al.}(2021)\citenamefont
		{Menahem}, \citenamefont {Dai}, \citenamefont {Aharon}, \citenamefont
		{Sharma}, \citenamefont {Asher}, \citenamefont {Diskin-Posner}, \citenamefont
		{Korobko}, \citenamefont {Rappe},\ and\ \citenamefont {Yaffe}}]{Menahem2021}%
	\BibitemOpen
	\bibfield  {author} {\bibinfo {author} {\bibfnamefont {M.}~\bibnamefont
			{Menahem}}, \bibinfo {author} {\bibfnamefont {Z.}~\bibnamefont {Dai}},
		\bibinfo {author} {\bibfnamefont {S.}~\bibnamefont {Aharon}}, \bibinfo
		{author} {\bibfnamefont {R.}~\bibnamefont {Sharma}}, \bibinfo {author}
		{\bibfnamefont {M.}~\bibnamefont {Asher}}, \bibinfo {author} {\bibfnamefont
			{Y.}~\bibnamefont {Diskin-Posner}}, \bibinfo {author} {\bibfnamefont
			{R.}~\bibnamefont {Korobko}}, \bibinfo {author} {\bibfnamefont {A.~M.}\
			\bibnamefont {Rappe}},\ and\ \bibinfo {author} {\bibfnamefont
			{O.}~\bibnamefont {Yaffe}},\ }\bibfield  {title} {\bibinfo {title} {{Strongly
				Anharmonic Octahedral Tilting in Two-Dimensional Hybrid Halide
				Perovskites}},\ }\href
	{https://doi.org/10.1021/ACSNANO.1C02022/SUPPL_FILE/NN1C02022_SI_001.PDF}
	{\bibfield  {journal} {\bibinfo  {journal} {ACS Nano}\ }\textbf {\bibinfo
			{volume} {15}},\ \bibinfo {pages} {10153} (\bibinfo {year}
		{2021})}\BibitemShut {NoStop}%
	\bibitem [{\citenamefont {He}\ \emph {et~al.}(2017)\citenamefont {He},
		\citenamefont {Zhu},\ and\ \citenamefont {Mo}}]{He2017}%
	\BibitemOpen
	\bibfield  {author} {\bibinfo {author} {\bibfnamefont {X.}~\bibnamefont
			{He}}, \bibinfo {author} {\bibfnamefont {Y.}~\bibnamefont {Zhu}},\ and\
		\bibinfo {author} {\bibfnamefont {Y.}~\bibnamefont {Mo}},\ }\bibfield
	{title} {\bibinfo {title} {{Origin of fast ion diffusion in super-ionic
				conductors}},\ }\href {https://doi.org/10.1038/ncomms15893} {\bibfield
		{journal} {\bibinfo  {journal} {Nat. Commun.}\ }\textbf {\bibinfo {volume}
			{8}},\ \bibinfo {pages} {1} (\bibinfo {year} {2017})}\BibitemShut {NoStop}%
	\bibitem [{\citenamefont {Dieterich}(1989)}]{Physik}%
	\BibitemOpen
	\bibfield  {author} {\bibinfo {author} {\bibfnamefont {W.}~\bibnamefont
			{Dieterich}},\ }\bibinfo {title} {Transport in ionic solids: Theoretical
		aspects},\ in\ \href {https://doi.org/10.1142/9789814434294_0002} {\emph
		{\bibinfo {booktitle} {High Conductivity Solid Ionic Conductors}}},\ \bibinfo
	{editor} {edited by\ \bibinfo {editor} {\bibfnamefont {T.}~\bibnamefont
			{Takahashi}}}\ (\bibinfo  {publisher} {World Scientific},\ \bibinfo {year}
	{1989})\ pp.\ \bibinfo {pages} {17--44}\BibitemShut {NoStop}%
	\bibitem [{\citenamefont {Brenner}\ \emph
		{et~al.}(2020{\natexlab{b}})\citenamefont {Brenner}, \citenamefont
		{Gehrmann}, \citenamefont {Korobko}, \citenamefont {Livneh}, \citenamefont
		{Egger},\ and\ \citenamefont {Yaffe}}]{PhysRevMaterials.4.115402}%
	\BibitemOpen
	\bibfield  {author} {\bibinfo {author} {\bibfnamefont {T.~M.}\ \bibnamefont
			{Brenner}}, \bibinfo {author} {\bibfnamefont {C.}~\bibnamefont {Gehrmann}},
		\bibinfo {author} {\bibfnamefont {R.}~\bibnamefont {Korobko}}, \bibinfo
		{author} {\bibfnamefont {T.}~\bibnamefont {Livneh}}, \bibinfo {author}
		{\bibfnamefont {D.~A.}\ \bibnamefont {Egger}},\ and\ \bibinfo {author}
		{\bibfnamefont {O.}~\bibnamefont {Yaffe}},\ }\bibfield  {title} {\bibinfo
		{title} {Anharmonic host-lattice dynamics enable fast ion conduction in
			superionic agi},\ }\href {https://doi.org/10.1103/PhysRevMaterials.4.115402}
	{\bibfield  {journal} {\bibinfo  {journal} {Phys. Rev. Materials}\ }\textbf
		{\bibinfo {volume} {4}},\ \bibinfo {pages} {115402} (\bibinfo {year}
		{2020}{\natexlab{b}})}\BibitemShut {NoStop}%
	\bibitem [{\citenamefont {Zhang}\ \emph {et~al.}(2019)\citenamefont {Zhang},
		\citenamefont {Roy}, \citenamefont {Li}, \citenamefont {Avdeev},\ and\
		\citenamefont {Nazar}}]{Zhang2019}%
	\BibitemOpen
	\bibfield  {author} {\bibinfo {author} {\bibfnamefont {Z.}~\bibnamefont
			{Zhang}}, \bibinfo {author} {\bibfnamefont {P.-N.}\ \bibnamefont {Roy}},
		\bibinfo {author} {\bibfnamefont {H.}~\bibnamefont {Li}}, \bibinfo {author}
		{\bibfnamefont {M.}~\bibnamefont {Avdeev}},\ and\ \bibinfo {author}
		{\bibfnamefont {L.~F.}\ \bibnamefont {Nazar}},\ }\bibfield  {title} {\bibinfo
		{title} {{Coupled Cation–Anion Dynamics Enhances Cation Mobility in
				Room-Temperature Superionic Solid-State Electrolytes}},\ }\href
	{https://doi.org/10.1021/jacs.9b09343} {\bibfield  {journal} {\bibinfo
			{journal} {J. Am. Chem. Soc.}\ }\textbf {\bibinfo {volume} {141}},\ \bibinfo
		{pages} {19360} (\bibinfo {year} {2019})}\BibitemShut {NoStop}%
	\bibitem [{\citenamefont {Muy}\ \emph {et~al.}(2018)\citenamefont {Muy},
		\citenamefont {Bachman}, \citenamefont {Giordano}, \citenamefont {Chang},
		\citenamefont {Abernathy}, \citenamefont {Bansal}, \citenamefont {Delaire},
		\citenamefont {Hori}, \citenamefont {Kanno}, \citenamefont {Maglia},
		\citenamefont {Lupart}, \citenamefont {Lamp},\ and\ \citenamefont
		{Shao-Horn}}]{C7EE03364H}%
	\BibitemOpen
	\bibfield  {author} {\bibinfo {author} {\bibfnamefont {S.}~\bibnamefont
			{Muy}}, \bibinfo {author} {\bibfnamefont {J.~C.}\ \bibnamefont {Bachman}},
		\bibinfo {author} {\bibfnamefont {L.}~\bibnamefont {Giordano}}, \bibinfo
		{author} {\bibfnamefont {H.-H.}\ \bibnamefont {Chang}}, \bibinfo {author}
		{\bibfnamefont {D.~L.}\ \bibnamefont {Abernathy}}, \bibinfo {author}
		{\bibfnamefont {D.}~\bibnamefont {Bansal}}, \bibinfo {author} {\bibfnamefont
			{O.}~\bibnamefont {Delaire}}, \bibinfo {author} {\bibfnamefont
			{S.}~\bibnamefont {Hori}}, \bibinfo {author} {\bibfnamefont {R.}~\bibnamefont
			{Kanno}}, \bibinfo {author} {\bibfnamefont {F.}~\bibnamefont {Maglia}},
		\bibinfo {author} {\bibfnamefont {S.}~\bibnamefont {Lupart}}, \bibinfo
		{author} {\bibfnamefont {P.}~\bibnamefont {Lamp}},\ and\ \bibinfo {author}
		{\bibfnamefont {Y.}~\bibnamefont {Shao-Horn}},\ }\bibfield  {title} {\bibinfo
		{title} {{Tuning mobility and stability of lithium ion conductors based on
				lattice dynamics}},\ }\href {https://doi.org/10.1039/C7EE03364H} {\bibfield
		{journal} {\bibinfo  {journal} {Energy Environ. Sci.}\ }\textbf {\bibinfo
			{volume} {11}},\ \bibinfo {pages} {850} (\bibinfo {year} {2018})}\BibitemShut
	{NoStop}%
	\bibitem [{\citenamefont {Krauskopf}\ \emph {et~al.}(2018)\citenamefont
		{Krauskopf}, \citenamefont {Muy}, \citenamefont {Culver}, \citenamefont
		{Ohno}, \citenamefont {Delaire}, \citenamefont {Shao-Horn},\ and\
		\citenamefont {Zeier}}]{Krauskopf2018}%
	\BibitemOpen
	\bibfield  {author} {\bibinfo {author} {\bibfnamefont {T.}~\bibnamefont
			{Krauskopf}}, \bibinfo {author} {\bibfnamefont {S.}~\bibnamefont {Muy}},
		\bibinfo {author} {\bibfnamefont {S.~P.}\ \bibnamefont {Culver}}, \bibinfo
		{author} {\bibfnamefont {S.}~\bibnamefont {Ohno}}, \bibinfo {author}
		{\bibfnamefont {O.}~\bibnamefont {Delaire}}, \bibinfo {author} {\bibfnamefont
			{Y.}~\bibnamefont {Shao-Horn}},\ and\ \bibinfo {author} {\bibfnamefont
			{W.~G.}\ \bibnamefont {Zeier}},\ }\bibfield  {title} {\bibinfo {title}
		{{Comparing the Descriptors for Investigating the Influence of Lattice
				Dynamics on Ionic Transport Using the Superionic Conductor Na3PS4–xSex}},\
	}\href {https://doi.org/10.1021/jacs.8b09340} {\bibfield  {journal} {\bibinfo
			{journal} {J. Am. Chem. Soc.}\ }\textbf {\bibinfo {volume} {140}},\ \bibinfo
		{pages} {14464} (\bibinfo {year} {2018})}\BibitemShut {NoStop}%
	\bibitem [{\citenamefont {Krauskopf}\ \emph {et~al.}(2017)\citenamefont
		{Krauskopf}, \citenamefont {Pompe}, \citenamefont {Kraft},\ and\
		\citenamefont {Zeier}}]{Krauskopf2017}%
	\BibitemOpen
	\bibfield  {author} {\bibinfo {author} {\bibfnamefont {T.}~\bibnamefont
			{Krauskopf}}, \bibinfo {author} {\bibfnamefont {C.}~\bibnamefont {Pompe}},
		\bibinfo {author} {\bibfnamefont {M.~A.}\ \bibnamefont {Kraft}},\ and\
		\bibinfo {author} {\bibfnamefont {W.~G.}\ \bibnamefont {Zeier}},\ }\bibfield
	{title} {\bibinfo {title} {{Influence of Lattice Dynamics on Na+ Transport in
				the Solid Electrolyte Na3PS4–xSex}},\ }\href
	{https://doi.org/10.1021/acs.chemmater.7b03474} {\bibfield  {journal}
		{\bibinfo  {journal} {Chem. Mater.}\ }\textbf {\bibinfo {volume} {29}},\
		\bibinfo {pages} {8859} (\bibinfo {year} {2017})}\BibitemShut {NoStop}%
	\bibitem [{\citenamefont {Kraft}\ \emph {et~al.}(2017)\citenamefont {Kraft},
		\citenamefont {Culver}, \citenamefont {Calderon}, \citenamefont
		{B{\"{o}}cher}, \citenamefont {Krauskopf}, \citenamefont {Senyshyn},
		\citenamefont {Dietrich}, \citenamefont {Zevalkink}, \citenamefont {Janek},\
		and\ \citenamefont {Zeier}}]{Kraft2017}%
	\BibitemOpen
	\bibfield  {author} {\bibinfo {author} {\bibfnamefont {M.~A.}\ \bibnamefont
			{Kraft}}, \bibinfo {author} {\bibfnamefont {S.~P.}\ \bibnamefont {Culver}},
		\bibinfo {author} {\bibfnamefont {M.}~\bibnamefont {Calderon}}, \bibinfo
		{author} {\bibfnamefont {F.}~\bibnamefont {B{\"{o}}cher}}, \bibinfo {author}
		{\bibfnamefont {T.}~\bibnamefont {Krauskopf}}, \bibinfo {author}
		{\bibfnamefont {A.}~\bibnamefont {Senyshyn}}, \bibinfo {author}
		{\bibfnamefont {C.}~\bibnamefont {Dietrich}}, \bibinfo {author}
		{\bibfnamefont {A.}~\bibnamefont {Zevalkink}}, \bibinfo {author}
		{\bibfnamefont {J.}~\bibnamefont {Janek}},\ and\ \bibinfo {author}
		{\bibfnamefont {W.~G.}\ \bibnamefont {Zeier}},\ }\bibfield  {title} {\bibinfo
		{title} {{Influence of Lattice Polarizability on the Ionic Conductivity in
				the Lithium Superionic Argyrodites Li6PS5X (X = Cl, Br, I)}},\ }\href
	{https://doi.org/10.1021/jacs.7b06327} {\bibfield  {journal} {\bibinfo
			{journal} {J. Am. Chem. Soc.}\ }\textbf {\bibinfo {volume} {139}},\ \bibinfo
		{pages} {10909} (\bibinfo {year} {2017})}\BibitemShut {NoStop}%
	\bibitem [{\citenamefont {Smith}\ and\ \citenamefont
		{Siegel}(2020)}]{Smith2020}%
	\BibitemOpen
	\bibfield  {author} {\bibinfo {author} {\bibfnamefont {J.~G.}\ \bibnamefont
			{Smith}}\ and\ \bibinfo {author} {\bibfnamefont {D.~J.}\ \bibnamefont
			{Siegel}},\ }\bibfield  {title} {\bibinfo {title} {{Low-temperature
				paddlewheel effect in glassy solid electrolytes}},\ }\href
	{https://doi.org/10.1038/s41467-020-15245-5} {\bibfield  {journal} {\bibinfo
			{journal} {Nat. Commun.}\ }\textbf {\bibinfo {volume} {11}},\ \bibinfo
		{pages} {1483} (\bibinfo {year} {2020})}\BibitemShut {NoStop}%
	\bibitem [{\citenamefont {Duch{\^{e}}ne}\ \emph {et~al.}(2019)\citenamefont
		{Duch{\^{e}}ne}, \citenamefont {Lunghammer}, \citenamefont {Burankova},
		\citenamefont {Liao}, \citenamefont {Embs}, \citenamefont {Cop{\'{e}}ret},
		\citenamefont {Wilkening}, \citenamefont {Remhof}, \citenamefont {Hagemann},\
		and\ \citenamefont {Battaglia}}]{Duchene2019}%
	\BibitemOpen
	\bibfield  {author} {\bibinfo {author} {\bibfnamefont {L.}~\bibnamefont
			{Duch{\^{e}}ne}}, \bibinfo {author} {\bibfnamefont {S.}~\bibnamefont
			{Lunghammer}}, \bibinfo {author} {\bibfnamefont {T.}~\bibnamefont
			{Burankova}}, \bibinfo {author} {\bibfnamefont {W.-C.}\ \bibnamefont {Liao}},
		\bibinfo {author} {\bibfnamefont {J.~P.}\ \bibnamefont {Embs}}, \bibinfo
		{author} {\bibfnamefont {C.}~\bibnamefont {Cop{\'{e}}ret}}, \bibinfo {author}
		{\bibfnamefont {H.~M.~R.}\ \bibnamefont {Wilkening}}, \bibinfo {author}
		{\bibfnamefont {A.}~\bibnamefont {Remhof}}, \bibinfo {author} {\bibfnamefont
			{H.}~\bibnamefont {Hagemann}},\ and\ \bibinfo {author} {\bibfnamefont
			{C.}~\bibnamefont {Battaglia}},\ }\bibfield  {title} {\bibinfo {title}
		{{Ionic Conduction Mechanism in the Na2(B12H12)0.5(B10H10)0.5 closo-Borate
				Solid-State Electrolyte: Interplay of Disorder and Ion–Ion Interactions}},\
	}\href {https://doi.org/10.1021/acs.chemmater.9b00610} {\bibfield  {journal}
		{\bibinfo  {journal} {Chem. Mater.}\ }\textbf {\bibinfo {volume} {31}},\
		\bibinfo {pages} {3449} (\bibinfo {year} {2019})}\BibitemShut {NoStop}%
	\bibitem [{\citenamefont {Adelstein}\ and\ \citenamefont
		{Wood}(2016)}]{Adelstein2016}%
	\BibitemOpen
	\bibfield  {author} {\bibinfo {author} {\bibfnamefont {N.}~\bibnamefont
			{Adelstein}}\ and\ \bibinfo {author} {\bibfnamefont {B.~C.}\ \bibnamefont
			{Wood}},\ }\bibfield  {title} {\bibinfo {title} {{Role of Dynamically
				Frustrated Bond Disorder in a Li+ Superionic Solid Electrolyte}},\ }\href
	{https://doi.org/10.1021/acs.chemmater.6b00790} {\bibfield  {journal}
		{\bibinfo  {journal} {Chem. Mater.}\ }\textbf {\bibinfo {volume} {28}},\
		\bibinfo {pages} {7218} (\bibinfo {year} {2016})}\BibitemShut {NoStop}%
	\bibitem [{\citenamefont {Kweon}\ \emph {et~al.}(2017)\citenamefont {Kweon},
		\citenamefont {Varley}, \citenamefont {Shea}, \citenamefont {Adelstein},
		\citenamefont {Mehta}, \citenamefont {Heo}, \citenamefont {Udovic},
		\citenamefont {Stavila},\ and\ \citenamefont {Wood}}]{Kweon2017}%
	\BibitemOpen
	\bibfield  {author} {\bibinfo {author} {\bibfnamefont {K.~E.}\ \bibnamefont
			{Kweon}}, \bibinfo {author} {\bibfnamefont {J.~B.}\ \bibnamefont {Varley}},
		\bibinfo {author} {\bibfnamefont {P.}~\bibnamefont {Shea}}, \bibinfo {author}
		{\bibfnamefont {N.}~\bibnamefont {Adelstein}}, \bibinfo {author}
		{\bibfnamefont {P.}~\bibnamefont {Mehta}}, \bibinfo {author} {\bibfnamefont
			{T.~W.}\ \bibnamefont {Heo}}, \bibinfo {author} {\bibfnamefont {T.~J.}\
			\bibnamefont {Udovic}}, \bibinfo {author} {\bibfnamefont {V.}~\bibnamefont
			{Stavila}},\ and\ \bibinfo {author} {\bibfnamefont {B.~C.}\ \bibnamefont
			{Wood}},\ }\bibfield  {title} {\bibinfo {title} {{Structural, Chemical, and
				Dynamical Frustration: Origins of Superionic Conductivity in closo-Borate
				Solid Electrolytes}},\ }\href {https://doi.org/10.1021/acs.chemmater.7b02902}
	{\bibfield  {journal} {\bibinfo  {journal} {Chem. Mater.}\ }\textbf {\bibinfo
			{volume} {29}},\ \bibinfo {pages} {9142} (\bibinfo {year}
		{2017})}\BibitemShut {NoStop}%
\end{thebibliography}

\begin{thebibliography}{2}%
	\makeatletter
	\providecommand \@ifxundefined [1]{%
		\@ifx{#1\undefined}
	}%
	\providecommand \@ifnum [1]{%
		\ifnum #1\expandafter \@firstoftwo
		\else \expandafter \@secondoftwo
		\fi
	}%
	\providecommand \@ifx [1]{%
		\ifx #1\expandafter \@firstoftwo
		\else \expandafter \@secondoftwo
		\fi
	}%
	\providecommand \natexlab [1]{#1}%
	\providecommand \enquote  [1]{``#1''}%
	\providecommand \bibnamefont  [1]{#1}%
	\providecommand \bibfnamefont [1]{#1}%
	\providecommand \citenamefont [1]{#1}%
	\providecommand \href@noop [0]{\@secondoftwo}%
	\providecommand \href [0]{\begingroup \@sanitize@url \@href}%
	\providecommand \@href[1]{\@@startlink{#1}\@@href}%
	\providecommand \@@href[1]{\endgroup#1\@@endlink}%
	\providecommand \@sanitize@url [0]{\catcode `\\12\catcode `\$12\catcode
		`\&12\catcode `\#12\catcode `\^12\catcode `\_12\catcode `\%12\relax}%
	\providecommand \@@startlink[1]{}%
	\providecommand \@@endlink[0]{}%
	\providecommand \url  [0]{\begingroup\@sanitize@url \@url }%
	\providecommand \@url [1]{\endgroup\@href {#1}{\urlprefix }}%
	\providecommand \urlprefix  [0]{URL }%
	\providecommand \Eprint [0]{\href }%
	\providecommand \doibase [0]{https://doi.org/}%
	\providecommand \selectlanguage [0]{\@gobble}%
	\providecommand \bibinfo  [0]{\@secondoftwo}%
	\providecommand \bibfield  [0]{\@secondoftwo}%
	\providecommand \translation [1]{[#1]}%
	\providecommand \BibitemOpen [0]{}%
	\providecommand \bibitemStop [0]{}%
	\providecommand \bibitemNoStop [0]{.\EOS\space}%
	\providecommand \EOS [0]{\spacefactor3000\relax}%
	\providecommand \BibitemShut  [1]{\csname bibitem#1\endcsname}%
	\let\auto@bib@innerbib\@empty
	\bibitem [{\citenamefont {Asher}\ \emph {et~al.}(2020)\citenamefont {Asher},
		\citenamefont {Angerer}, \citenamefont {Korobko}, \citenamefont
		{Diskin-Posner}, \citenamefont {Egger},\ and\ \citenamefont
		{Yaffe}}]{Asher20p1908028}%
	\BibitemOpen
	\bibfield  {author} {\bibinfo {author} {\bibfnamefont {M.}~\bibnamefont
			{Asher}}, \bibinfo {author} {\bibfnamefont {D.}~\bibnamefont {Angerer}},
		\bibinfo {author} {\bibfnamefont {R.}~\bibnamefont {Korobko}}, \bibinfo
		{author} {\bibfnamefont {Y.}~\bibnamefont {Diskin-Posner}}, \bibinfo {author}
		{\bibfnamefont {D.~A.}\ \bibnamefont {Egger}},\ and\ \bibinfo {author}
		{\bibfnamefont {O.}~\bibnamefont {Yaffe}},\ }\bibfield  {title} {\bibinfo
		{title} {{Anharmonic Lattice Vibrations in Small-Molecule Organic
				Semiconductors}},\ }\href
	{https://doi.org/https://doi.org/10.1002/adma.201908028} {\bibfield
		{journal} {\bibinfo  {journal} {Adv. Mater.}\ }\textbf {\bibinfo {volume}
			{32}},\ \bibinfo {pages} {1908028} (\bibinfo {year} {2020})}\BibitemShut
	{NoStop}%
	\bibitem [{\citenamefont {Sharma}\ \emph {et~al.}(2020)\citenamefont {Sharma},
		\citenamefont {Dai}, \citenamefont {Gao}, \citenamefont {Brenner},
		\citenamefont {Yadgarov}, \citenamefont {Zhang}, \citenamefont {Rakita},
		\citenamefont {Korobko}, \citenamefont {Rappe},\ and\ \citenamefont
		{Yaffe}}]{Sharma}%
	\BibitemOpen
	\bibfield  {author} {\bibinfo {author} {\bibfnamefont {R.}~\bibnamefont
			{Sharma}}, \bibinfo {author} {\bibfnamefont {Z.}~\bibnamefont {Dai}},
		\bibinfo {author} {\bibfnamefont {L.}~\bibnamefont {Gao}}, \bibinfo {author}
		{\bibfnamefont {T.~M.}\ \bibnamefont {Brenner}}, \bibinfo {author}
		{\bibfnamefont {L.}~\bibnamefont {Yadgarov}}, \bibinfo {author}
		{\bibfnamefont {J.}~\bibnamefont {Zhang}}, \bibinfo {author} {\bibfnamefont
			{Y.}~\bibnamefont {Rakita}}, \bibinfo {author} {\bibfnamefont
			{R.}~\bibnamefont {Korobko}}, \bibinfo {author} {\bibfnamefont {A.~M.}\
			\bibnamefont {Rappe}},\ and\ \bibinfo {author} {\bibfnamefont
			{O.}~\bibnamefont {Yaffe}},\ }\bibfield  {title} {\bibinfo {title}
		{Elucidating the atomistic origin of anharmonicity in tetragonal
			${\mathrm{ch}}_{3}{\mathrm{nh}}_{3}{\mathrm{pbi}}_{3}$ with raman
			scattering},\ }\href {https://doi.org/10.1103/PhysRevMaterials.4.092401}
	{\bibfield  {journal} {\bibinfo  {journal} {Phys. Rev. Mater.}\ }\textbf
		{\bibinfo {volume} {4}},\ \bibinfo {pages} {092401} (\bibinfo {year}
		{2020})}\BibitemShut {NoStop}%
\end{thebibliography}
\end{document}